
%

\magnification=\magstep1
\spaceskip=0.4em plus 0.15em minus 0.15em
\xspaceskip=0.5em
\hsize=16 true cm
\hoffset=0.25 true cm
\vsize=22 true cm
\hyphenpenalty=10000
\tolerance=1000
\headline={\hss\tenrm\folio}
\footline={\hss}


\def\dsp{\baselineskip=22pt plus 1pt minus 1pt}


.300
.518 at 14.4pt
\font\chapterfont=cmb10.518 at 14.4pt
\font\sectionfont=cmb10.328 at 12pt
\font\subsectionfont=cmbx10.300

\newcount\sectno
\newcount\subsectno

\def\section#1{\global\advance\sectno by 1 \subsectno=0
 \message{#1}
 \removelastskip\vskip 20pt plus 50pt\penalty-200
 \vskip 0pt plus -42pt\centerline{\sectionfont\the\sectno.\ #1}
 \nobreak\medskip
}
\def\sectiont#1#2{\global\advance\sectno by 1 \subsectno=0
 \message{#1 #2}
 \removelastskip\vskip 20pt plus 50pt\penalty-200
 \vskip 0pt plus -42pt\centerline{\sectionfont\the\sectno.\ #1}
 \smallskip\centerline{\sectionfont #2}
 \nobreak\medskip
}

\def\subsection#1{\global\advance\subsectno by 1
 \medbreak\vskip10pt
 \leftline{\subsectionfont\the\sectno.\the\subsectno.\ #1}
 \nobreak\smallskip
}
\def\subsectiont#1#2{\global\advance\subsectno by 1
 \medbreak\vskip10pt
 \leftline{\subsectionfont\the\sectno.\the\subsectno.\ #1}
 \leftline{\hskip1.3truecm\subsectionfont #2}
 \nobreak\smallskip
}

\def\appendix#1#2{\vfil\eject
 \message{Appendix #1} \null\vskip.5truecm
 \centerline{\sectionfont APPENDIX #1. #2}
 \nobreak\medskip
}


\def\widedotfill{\leaders\hbox to 15pt{\hfil.\hfil}\hfill}
\def\page#1{\widedotfill\rlap{\hbox to 25pt{\hfil#1}}\par}

\def\figures{\vfill\eject \centerline{FIGURE CAPTIONS} \bigskip}
\def\fig{\message{Figure Captions}
 \medskip\hangindent=2truecm \hangafter=1 \noindent}









\def\title#1{\dsp\centerline {\uppercase{#1}}}
\def\author#1{\bigskip\medskip \centerline {\uppercase{#1}}}

\def\and{\bigskip \centerline{and}}


\def\abstract{\vfil\eject\message{Abstract}
 \null\vskip0truecm\leftline{\hbox{\chapterfont Abstract}}
 \vskip.5truecm}
\def\subject#1{\bigskip\noindent SUBJECT HEADINGS:\ \ #1}

\def\ack{\bigskip\bigskip \goodbreak \indent}



\def\rhang{\dsp\noindent\hangindent=0.25truein\hangafter1\rm}
\def\references{\vfil\eject\message{References}
 \null\vskip0truecm\leftline{\hbox{\chapterfont References}}
 \vskip2.truecm\parskip=11pt}

\def\apj{ApJ}

\def\apjs{ApJ~Supp}

\def\mnras{MNRAS}
\def\aa{A\&A}
\def\aas{A\&A~Supp}

\def\rj#1#2#3#4{\rhang#1, #2, #3, #4\par}

\def\rb#1#2#3{\rhang#1, #2, \rm\ #3\par}

\def\rc#1#2#3#4#5{\rhang#1, in #2, ed. #3, #4, p.~#5\par}
\def\lsim{\lower2pt\hbox{$\buildrel {\scriptstyle <}
   \over {\scriptstyle\sim}$}}

\def\gsim{\lower2pt\hbox{$\buildrel {\scriptstyle >}
   \over {\scriptstyle\sim}$}}

\def\lapprox{\lower2pt\hbox{$\buildrel \lower2pt\hbox{${\scriptstyle<}$}
   \over {\scriptstyle\approx}$}}

\def\gapprox{\lower2pt\hbox{$\buildrel \lower2pt\hbox{${\scriptstyle>}$}
   \over {\scriptstyle\approx}$}}

\def\bbuildrel#1_#2^#3{\mathrel{
 \mathop{\kern 0pt#1}\limits_{#2}^{#3}}}

\def\etal{{et~al.\ }}
\def\eg{{e.g.}}
\def\vs{{vs.\ }}
\def\ie{{i.e.}}

\def\cf{{cf.\ }}

\def\km{\,\rm km}

\def\rhob{\overline\rho}
\def\vb{\overline v}
\def\Tb{\overline T}

\def\Kb{\overline K}
\def\Qdb{\overline{\dot Q}}

\def\cp{c_P}
\def\kb{k_{\rm B}}
\def\numic{\nu_{\rm mic}}

\def\hp{H_P}
\def\wwb{\overline{w^2}}
\def\ttb{\overline{\theta^2}}
\def\wtb{\overline{w\theta}}

\def\uub{\overline{u^2}}
\def\utb{\overline{u\theta}}
\def\wub{\overline{wu}}
\def\wwwb{\overline{w^3}}
\def\wwtb{\overline{w^2\theta}}
\def\wttb{\overline{w\theta^2}}
\def\tttb{\overline{\theta^3}}

\def\uuub{\overline{u^3}}
\def\wuub{\overline{wu^2}}
\def\wwub{\overline{w^2u}}
\def\uutb{\overline{u^2\theta}}
\def\uttb{\overline{u\theta^2}}
\def\wwwwb{\overline{w^4}}
\def\wwwtb{\overline{w^3\theta}}
\def\wwttb{\overline{w^2\theta^2}}
\def\wtttb{\overline{w\theta^3}}
\def\ttttb{\overline{\theta^4}}
\def\uuuub{\overline{u^4}}
\def\wwuub{\overline{w^2u^2}}
\def\wuutb{\overline{wu^2\theta}}
\def\uuttb{\overline{u^2\theta^2}}

\def\partt{\partial t}
\def\partz{\partial z}
\def\popz{{\partial\over{\partial z}}}
\def\pvopz{{\partial\vb_z\over{\partial z}}}
\def\pPopz{{\partial P\over{\partial z}}}
\def\dgt{\Delta\nabla T}

\def\lh{\ell_{\rm H}}
\def\lv{\ell_{\rm V}}

\def\fa{f_{\rm A}}
\def\fr{f_{\rm R}}

\def\Fconv{F_{\rm conv}}

\def\hp{H_{\rm P}}

\def\dover{d_{\rm over}}

\hfuzz15pt

\null\vskip75pt
\centerline{\bf A Theory of Non-Local Mixing-Length}
\centerline{\bf Convection.  III. Comparing Theory }
\centerline{\bf and Numerical Experiment}
\medskip
\centerline{Scott~A.~Grossman$^1$}
\medskip\centerline{MNRAS in press}
\vskip20pt
\centerline{Canadian Institute for Theoretical Astrophysics}
\centerline{60 St. George Street}
\centerline{Toronto, Ontario M5S 1A7}
\centerline{Canada}

\vskip.1truein
\hrule width 120pt
{\baselineskip12pt \noindent $^1$Current address: Northwestern
University, Dearborn Observatory, 2131 Sheridan Rd., Evanston, IL
60208}

\def\abstract{\message{Abstract}
 \null\vskip0truecm\leftline{\hbox{\chapterfont Abstract}}
 \vskip.5truecm}
\abstract We solve the nonlocal convection equations.  The solutions
for four model problems are compared with results of GSPH simulations.
In each case we test two closure schemes: 1) where third moments are
defined by the diffusion approximation; and 2) where the full third
moment equations are used and fourth moments are defined by a modified
form of the quasi-normal approximation.  In overshooting models, the
convective flux becomes negative shortly after the stability boundary.
The negative amplitude remains small, and the temperature gradient in
the overshooting zone has nearly the radiative value.  Turbulent
velocities decay by a factor of $e$ after $0.5$--$1.5\ell$, depending
on the model, where $\ell$ is the mixing length.  Turbulent viscosity
is more important than negative buoyancy in decelerating overshooting
fluid blobs.  These predictions are consistent with helioseismology.

The equations and the GSPH code use the same physical approximations,
so it was anticipated that if the closures for high order moments are
accurate enough, solutions for the low order moments will
automatically agree with the GSPH results.  Such internal consistency
holds approximately.  Unexpectedly, however, the best second moments
are found with the first closure scheme, and the best third moments
are obtained with the second.  The relationship among moments from the
solution of the equations is not the same as the relationship found by
GSPH simulation.  In particular, if we use the best fourth moment
closure model suggested by Paper II, we cannot get steady state
solutions, but adding a sort of diffusion for stability helps.

\subject{convection--hydrodynamics--stars: interiors--turbulence}

\section{Introduction}

The turbulent mixing of material in a convective region into an
adjacent region of stability is called convective overshooting.  In
the broadest terms, convection and convective overshooting have two
major consequences for stars:  1) Convection makes the temperature
gradient in a star shallower than the radiative value, and usually
very close to the adiabatic gradient.  Convective overshooting also
may modify the temperature gradient in the overshoot region, with
implications for the hydrostatic structure of stars.  There continues
to be disagreement in the literature whether the overshooting zone is
nearly radiative or nearly adiabatic.  2) The composition in
convective and overshooting regions is homogenized.  In stars with
convective cores, the lifetimes of the various nuclear burning phases
can be altered significantly by overshooting, and in stars with
convective envelopes, undershooting that reaches into the interior can
modify the surface composition.  The consequences for stellar
evolution if overshooting is significant (say at least a few tenths of
a pressure scale height) have been investigated extensively (\eg,
Bertelli, Bressan, \& Chiosi 1985; Maeder \& Meynet 1989).
Nevertheless, whether or not overshooting is, in fact, important
remains uncertain.

Papers addressing convective overshooting have been written for more
than three decades, and still there are major qualitative
disagreements among authors regarding the extent and importance of
overshooting.  Several authors have concluded that overshooting is
significant (Shaviv \& Salpeter 1973; Maeder 1975; Bressan, Bertelli,
\& Chiosi 1981; Zahn 1991; Xiong \& Chen 1992; Roxburgh 1978, 1989,
1992 are some examples), whereas others have come to the opposite
conclusion (Travis \& Matsushima 1973; Langer 1986).  The difficulty
in resolving this issue theoretically is that convective overshooting
is a completely nonlocal problem.  That is, the behavior of the
overshooting region depends sensitively on the behavior of the
adjacent convective region.  The mixing-length theory (B\"ohm-Vitense
1958) usually used to describe stellar convection is strictly local.
It is not a simple matter to decide which authors are more likely to
be correct.  In fact, Renzini (1987) has criticized most of the above
works, on both sides of this issue, for internal inconsistencies and
unjustified assumptions.

As a result of theoretical uncertainties, most stellar models are
computed using local convection and no overshooting.  When overshooting
is included (\eg, Doom 1982a, 1982b; Maeder \& Meynet 1988; Chin \&
Stothers 1991), it usually is included using an overshooting distance
parametrized by $\alpha_{\rm over}=\dover/\hp$, the ratio of the
overshooting distance to the pressure scale height at the
Schwarzschild stability boundary.

Since a variety of observational consequences of overshooting are
known (see Stothers 1991 for a summary of many observational tests),
it is possible, in principle, to calibrate the overshooting parameter,
without an understanding of the physics of nonlocal convection.  Many
authors have attempted such a calibration, but as with the theoretical
work, there is disagreement among authors here also.  Several
comparisons between cluster data and theoretical H-R diagrams favor an
intermediate amount of overshooting ($\dover=0.2$--$0.3\hp$) from
stellar cores (Maeder \& Mermilliod 1981; Mermilliod \& Maeder 1986;
Chiosi \etal 1989), and undershooting from convective envelopes may be
required also (Alongi \etal 1991).  Other authors conclude that
stellar evolution can be understood without overshooting (Stothers
1991; Stothers \& Chin 1992) and set an upper limit of
$\dover<0.2\hp$.  Even if the empirical calibrations eventually
converge to a widely accepted value appropriate to certain regions of
certain kinds of stars, overshooting distances probably would depend
on the particular conditions of the star and probably would not apply
to both core and envelop convection in all types of stars .  Thus, a
theory for nonlocal convection ultimately will be required.

This paper is the third in a series devoted to the development of a
theory of nonlocal convection.  In Paper I of this series (Grossman,
Narayan, \& Arnett 1993), we developed a Boltzmann transport theory
for the evolution of turbulent fluid elements and derived the
equations for the hydrodynamic evolution of high order correlations of
velocity and temperature.  In that work, the state variables of each
fluid blob were its vertical position $z$, vertical velocity $v$, and
temperature $T$.  The ensemble of fluid blobs was described by the
absolute distribution function $\fa(t,z,v,T)$ or, equivalently, by the
relative distribution function $\fr(t,z,w,\theta)$ of perturbations
$w=v-\vb$, $\theta=T-\Tb$.  (Bars over variables indicate ensemble
averages.)  The distribution function evolved according to a
Boltzmann-like equation, given by
$${\partial\fr\over\partt}+\popz[(\vb+w)\fr] +{\partial\over{\partial
w}}(\dot w\fr)+{\partial\over{\partial\theta}}(\dot\theta\fr)
=\Gamma,\eqno(1)$$ where $\dot w$ and $\dot\theta$ represent the
dynamical equations for the evolution of a single fluid blob and
$\Gamma$ is a collision term.  Equation (1) connects convection
theories of the ballistic particle type (theories that integrate $\dot
w$ and $\dot\theta$) to those that use a hydrodynamic approach.  The
derivation assumes that fluid blobs are in pressure balance with the
local mean fluid.

In Paper I, the effects of turbulent viscosity and turbulent diffusion
entered the $\dot w$ and $\dot\theta$ equations as eddy-damping terms.
We assumed turbulent fluid blobs travel the characteristic distance of
a mixing length $\ell$ with characteristic turbulent velocity
$\sigma_w$ before giving up their excess momentum and heat to the
ambient fluid.  Hence, turbulent viscosity and diffusion coefficients
are defined as $\nu_{\rm turb}\sim\chi_{\rm turb}\sim\ell\sigma_w$.
Since we account for turbulent losses on the left-hand side of the
transport equation (1), we set the collision term on the right-hand
side to zero.  By taking $w$ and $\theta$ moments of the transport
equation, we derived all moment equations up to third order.  In an
alternative formulation of this problem, presented in detail in
Appendix B, we treat the effects of turbulent damping in the collision
term instead of the dynamical equations for $\dot w$ and $\dot\theta$.

The moment equations form an unclosed hierarchy, and thus require
closure relations if they are to be solved.  In Paper II of this
series (Grossman \& Narayan 1993), we simulated nonlocal convection in
a one-dimensional box using an algorithm we call Generalized Smoothed
Particle Hydrodynamics (GSPH).  This code used the same equations for
the evolution of velocities and temperatures, $\dot w$ and
$\dot\theta$, as the moment theory, thereby simulating the same
physics.  We presented results of four simulations.  Two were for
``homogeneous'' convection, that is, for fluids convective throughout,
one of which was in the regime of efficient convection and the other
in the regime of inefficient convection.  Two were overshooting
simulations, where fluids made the transition from instability to
stability at the center of the box.  The unstable region in one case
was in the regime of efficient convection and in the other in the
regime of inefficient convection.  We investigated the nonlocal
behavior of convecting fluids, emphasizing the physics of overshooting
regions.  The relationships between third and fourth moments to lower
order moments were studied, in order to discover useful closure
approximations.

In this paper we solve the moment equations of Paper I using closure
relations suggested by the GSPH simulations of Paper II.  We regard
the GSPH results as data to be modeled by a correct analytic
description of nonlocal convection.  Hence, we compare the
steady-state solutions of the moment equations to the results of the
four GSPH simulations presented in Paper II.

The equations are solved using two different closure schemes.  The
first, used previously in the astrophysical literature by Xiong (1980,
1981, 1989), closes the equations at the third moments using the
diffusion approximation.  Xiong's equations have been solved in detail
for the convection and overshooting zones of the Sun (Unno, Kondo, \&
Xiong 1985; Unno \& Kondo 1989; Xiong \& Chen 1992) and other stars of
various sorts (Xiong 1985, 1986, 1990).

The second closure scheme closes the equations at the fourth moments
using a modified version of the quasi-normal approximation.  To our
knowledge, hydrodynamic equations closed at this high order have never
before been solved in the astrophysical literature.  Canuto (1992,
1993) wrote equations to the same high order using the quasi-normal
closure scheme.  They have been solved for boundary conditions
appropriate for Earth's convective boundary layer (Canuto
\etal 1994).  One key difference between this problem and
astrophysical convection is that planetary convection can be described
as an initial value problem, and the equations can be integrated
outward.  The convective and total fluxes are computed in this
integration.  In stellar convection, the total flux is specified, and
boundary conditions must be placed on both the inner and outer
boundaries of the fluid.  Thus, solving the astrophysical problem is
more complicated.  Also, astrophysical convection occurs in zones many
pressure scale heights deep, whereas in the Earth convection occurs
over a fraction of a scale height (in a region $\sim 1\km$ thick
compared to a scale height $\sim 7\km$) and is driven not by the
superadiabatic temperature gradient, but by the temperature gradient.
Nevertheless, the favorable agreement between the computed moments and
their observed behavior in this context lends support to our approach
to nonlocal convection.

In this paper, we consider not only two different closure schemes for
our set of moment equations, but, in fact, two different sets of
moment equations.  The first set are simply the equations of Paper I.
The second set is derived using the alternative collisional treatment
of turbulent losses.  The differences in the equations seem minor, but
solutions show significant quantitative differences.  The latter
theory is somewhat flexible and may ultimately prove to be a more
useful set of equations.

The plan of the paper is as follows.  In \S2 we discuss the moment
equations and the closure relations that we solve.  We analyze the
moment equations to make rough analytic estimates of overshooting
distances.  These predictions are compared with those of Zahn (1991)
and with results inferred from helioseismology.  In \S3 we outline the
numerical method used to solve the moment equations.  The solutions
for our four models are presented in \S4, where we make a detailed
comparison with the GSPH results.  In \S5 we discuss the alternative
formulation of the moment equations, where turbulent losses originate
as scattering by the Boltzmann collision term.  We refer to these
alternate equations as the scattering equations, in which we also have
included turbulence in the horizontal dimensions.  We discuss the
closure relations used in the solution of these equations.  In \S6 the
equations are solved and compared to results of new GSPH simulations.
A discussion of the comparisons between theory and numerical
simulation and a conclusion appear in \S7.

\section{Paper I Equations}

\subsection{The Equations and Closures}

In Paper I we derived equations for the time evolution of the mean
density $\rhob$, mean vertical velocity $\vb$, and mean temperature
$\Tb$ of a convecting fluid.  In addition, we wrote evolution
equations for correlations of $w$ and $\theta$, the velocity and
temperature perturbations.  A minimal convection theory requires at
least the three second moment equations for $\wwb$, $\wtb$, and
$\ttb$.  The equations include gradients of third moments.  If these
third moment terms are dropped, standard local mixing length theory is
recovered, but to describe nonlocal convection, these third moments
must be kept.  They can be defined either by nontrivial closure
relations (unlike the trivial relations equating them to zero) or by
the four third moment equations for $\wwwb$, $\wwtb$, $\wttb$, and
$\tttb$.  In the latter case, the third moment equations include
gradients of fourth moments, for which we will require nontrivial
closures.

The two nontrivial closure schemes, to which we refer as the ``Xiong
solution'' and the ``Full solution,'' are outlined below.  The
equations that we solve here are slightly simplified from those of
Paper I and are presented in Appendix A.  They are supplemented by an
ideal gas equation of state.

\noindent{\it Xiong Solution}--A common third moment
closure is the diffusion or down-gradient approximation, which relates
third moments to gradients of second moments.  Although not invented
by Xiong, they have been advanced in the astrophysical literature by
him (Xiong 1980, 1981).  They can be written as $$\eqalignno{
\wwwb&=-c_1\ell\sigma_w\nabla\wwb,&(2a)\cr
\wwtb&=-c_2\ell\sigma_w\nabla\wtb,&(2b)\cr
\wttb&=-c_3\ell\sigma_w\nabla\ttb,&(2c)\cr}$$ where $\ell$ is the
mixing length and $\sigma_w$ is the turbulent velocity dispersion.
Averaged over all four simulations presented in Paper II, the
constants $c_1$, $c_2$, and $c_3$ were calibrated to have values
between 0.5 and 1.8.  These optimal values, however, had dispersions
from one simulation to the next of order 100\%.  In the solutions
below we have taken these constants equal to unity since the deviation
from the optimal values apparently makes little difference in the
second moments.  Of course, the third moments are affected more
directly by the values of these constants.  Although a closure for
$\tttb$ is not required, we, nevertheless, compute it according to
$\tttb=\wwtb\,\ttb/\wwb$ for comparison with the GSPH results.

\noindent{\it Full Solution}--The Full solution includes the
third moment equations, which are closed at the level of the fourth
moments.  The closures can be written as $$\eqalignno{
\wwwwb&=\zeta\wwb^2-d\ell\sigma_w\nabla\wwwb&(3a)\cr
\wwwtb&=\zeta\wwb\,\wtb-d\ell\sigma_w\nabla\wwtb&(3b)\cr
\wwttb&=\zeta({1\over 3}\wwb\,\ttb+{2\over 3}\wtb^2)
-d\ell\sigma_w\nabla\wttb&(3c)\cr
\wtttb&=\zeta\wtb\,\ttb-d\ell\sigma_w\nabla\tttb.&(3d)\cr}$$ Setting
$\zeta=3$ in each of the first terms and dropping the second terms
reproduce the standard quasi-normal approximation of mathematical
hydrodynamics (\eg, Lesieur 1987; Orszag 1977).  The second terms in
these closures cause diffusion of third moments upon substitution into
the third moment equations.  The parameter $d$ sets the diffusion
rate.  For most of our convection models (described below), we cannot
obtain solutions without this term.  If we set $d=0$, physically
positive quantities, such as $\wwb$ or $\ttb$, sometimes become
negative as solutions evolve toward the steady state.  It is likely
that the final $\wwb$ and $\ttb$ would be negative if it could reach a
steady state, and that these unphysical solutions are not merely a
transient state.  If this happens, our code eventually will fail to
converge and will not reach a steady state.  We find that using
$d=0.8$ allows steady state solutions of acceptable qualitative
agreement to our four standard problems.  We do not present any
results for a pure quasi-normal closure model.  (In the few cases
where we can obtain solutions with a pure quasi-normal closure, the
solutions compare with the GSPH simulations much more poorly.)  Note
that a closure for $\ttttb$ is not required.

\subsection{The Scale of Overshooting}

Before solving the moment equations numerically, we make some
estimates of the extent of convective overshooting based on analysis
of the equations.  To keep the problem tractable, we consider only the
Xiong closure here.  We simplify the second moment equations to a
point that captures only the essential nonlocal physics.  We consider
a quasi-homogeneous fluid of constant density $\rhob$ and temperature
$\Tb$, but with a variable superadiabatic gradient $\dgt$.  We assume
the turbulent velocity is sufficiently subsonic that gravity $g$ is
balanced entirely by the thermal pressure gradient and drop terms
associated with viscous heating, since they are always small in our
models.  The steady-state second moment equations from Appendix A
become $$\eqalignno{
{\partial\wwwb\over\partz}&-{2g\alpha\over\Tb}\wtb+2(A+B\sigma_w)\wwb=0,
&(4a)\cr {\partial\wwtb\over\partz}&-{g\alpha\over\Tb}\ttb
+2(A+B\sigma_w+D+E\sigma_w)\wtb-\dgt\wwb=0,&(4b)\cr
{\partial\wttb\over\partz}&-2\dgt\wtb+2(D+E\sigma_w)\ttb=0.&(4c)\cr}$$
The constants are defined in equations (A11), but briefly, $A$ and $D$
measure microscopic viscosity and radiative diffusion damping rates of
momentum and heat excesses, and $B\sigma_w$ and $E\sigma_w$ are the
damping rates by turbulent processes.  The constant $\alpha$ is the
coefficient of thermal expansion, equal to unity for an ideal gas.
The leading gradient terms are responsible for nonlocality; without
them, these equations would describe local mixing length theory.

Let us consider a fluid with an unstable superadiabatic gradient
$\dgt_1>0$ adjacent to a stable region with $\dgt_2<0$.  In the
unstable region, the second moments can be approximated by the values
of the local theory, so that at the stability boundary the second
moments can be written in terms of $\dgt_1$.  Convective overshooting
is a nonlocal process, and we estimate the magnitudes of the nonlocal
terms in the overshooting zone as the values needed to offset the
change in the superadiabatic gradient across the stability boundary.
If we describe the decay of second moments in the overshooting zone
with exponential dependences $\wwb\sim e^{-z/H_{w^2}}$,\ $\wtb\sim
e^{-z/H_{w\theta}}$, and $\ttb\sim e^{-z/H_{\theta^2}}$ ($z=0$
corresponds to the stability transition) and substitute the Xiong
closures into equations (4b) and (4c), we find for the decay length
scales
$$H_{w\theta}=H_{\theta^2}=\left({\ell\sigma_w\over{2(D+E\sigma_w)}}
{\dgt_1\over{\dgt_1-\dgt_2}}\right)^{1/2}.\eqno(5)$$ The closer
$\dgt_1$ is to zero, the shorter is the decay scale of moments
involving $\theta$.  The reason is that turbulent blobs moving
adiabatically accumulate less temperature excess as they move in the
unstable region, and hence lose what little they have accumulated more
quickly in the stable region.  If the Peclet number, the ratio of
turbulent to diffusive damping rates, $E\sigma_w/D\gg 1$, convection
is efficient and $$H_{w\theta}=H_{\theta^2}=\left({\ell\over 2}\right)
\left({\dgt_1\over{\dgt_1-\dgt_2}}\right)^{1/2}.\eqno(6)$$ If
$E\sigma_w/D\ll 1$, convection is inefficient and
$$H_{w\theta}=H_{\theta^2}=\left({E\sigma_w\over D}\right)^{1/2}
\left({\ell\over 2}\right)
\left({\dgt_1\over{\dgt_1-\dgt_2}}\right)^{1/2}.\eqno(7)$$

Because the superadiabatic gradient does not enter into equation (4a),
$\wwb$ does not respond immediately to the change in the
superadiabatic gradient.  As we just showed, $\wtb$ decays and an
overshooting particle loses its buoyancy.  When $\wtb$ becomes small,
the nonlocal term of equation (4a) must be balanced by the viscous
damping term, so that
$$H_{w^2}=\left({\ell\sigma_w\over{2(A+B\sigma_w)}}\right)^{1/2}.\eqno(8)$$
In all cases of astrophysical relevance, the turbulent viscosity far
exceeds the molecular viscosity, $B\sigma_w/A\gg 1$, so that
$$H_{w^2}=\left({\ell\over 2}\right).\eqno(9)$$ A typical overshooting
scale might be estimated as the sum $H_{w\theta}+H_{w^2}$.  Of course,
since the equations are nonlinear, the decays are not simple
exponentials, and these estimates are likely to be fairly crude.

To evaluate the usefulness of these quantitative estimates, in Figure
1 we show the local and nonlocal nonlinear solutions for two
overshooting models.  The regions of stability and instability are
each characterized by approximately constant superadiabatic gradients.
(The parameters of these models are the same as the overshooting
models discussed below, except that the diffusion coefficient is
constant in the upper and lower halves of the box and is discontinuous
across the stability boundary.)  The efficient overshooting model has
a convective zone in the efficient regime.  The solution to the local
equations is shown as dotted lines.  For the mixing length
$\ell=0.24$, we estimate $H_{w^2}\approx 0.12$, $H_{w\theta}\approx
0.07$, and $H_{\theta^2}\approx 0.07$, which compare favorably with
the distances $0.17$, $0.05$, and $0.05$ for the second moments to
fall by a factor of $e$ beyond the stability transition.  We note also
that the local and nonlocal solutions are comparable at the stability
transition, justifying our approximation that the second moments decay
from the local values at the stability boundary.

The inefficient overshooting model has a convective region in the
inefficient regime.  We estimate overshooting scales $H_{w^2}\approx
0.12$, $H_{w\theta}\approx 0.07$, and $H_{\theta^2}\approx 0.07$,
which compare well with the values $0.14$, $0.03$, and $0.03$ for the
real nonlinear solutions.  In the convective region, the difference
between the local and nonlocal solutions is greater in this case
because the convective flux is not constrained to be nearly the total
flux.  We note that the distance $H_{w^2}$ slightly underestimates the
actual overshooting distance of $\wwb$, whereas $H_{w^2}+H_{w\theta}$
slightly overestimates it.

\subsection{Zahn's Analysis}

We examine the analysis of Zahn (1991) within the framework of our
moment theory and consider whether his results are consistent with
ours.  From the $\Tb$ equation (A3), if there is no local heating or
cooling and a constant flux flows through a fluid, changes in the
radiative flux must be balanced by changes in the convective flux,
$\nabla(\Kb\nabla\Tb)=\nabla\rhob\cp\wtb$.  Zahn assumes the
overshooting region is nearly adiabatic, so that $\nabla\Tb$ is
approximately the constant, adiabatic value, $\nabla\Tb_{\rm ad}$.  He
also assumes that the outgoing convective flux falls to zero and then
changes sign at the beginning of the overshoot region.  Then,
expanding to linear order, he obtains for the convective flux
$$\Fconv=\rhob\cp\wtb\approx\nabla\Kb\nabla\Tb_{\rm ad}z,\eqno(10)$$
which is negative since the temperature gradient is negative.  We
think this is an inconsistency.  The convective flux does indeed
change sign slightly beyond the stability transition (see below), but
it does this precisely because overshooting blobs are moving against a
stable temperature gradient.  The convective flux cannot change sign
if the temperature gradient remains adiabatic.  Further, we find below
that when it does change sign, it does not obtain the large negative
values required to maintain a nearly adiabatic temperature gradient,
but remains relatively small.

According to Paper I (eq. 5.1), the velocity of the overshooting blob
evolves according to $$\dot w=w{\partial w\over\partz}=
{g\alpha\theta\over\Tb}-(A+B\sigma_w)w,\eqno(11)$$ which has
contributions due to both buoyancy and viscosity.  A large negative
convective flux implies a large negative buoyancy, which Zahn regards
as the primary reason overshooting blobs decelerate, and hence he
neglects the viscosity term.  Then, multiplying equation (11) by $w$,
using the result of equation (10), and integrating the velocity until
it decreases to zero, he obtains the overshooting distance $$z_{\rm
over}\approx (w_0)^{3/2}\left({3\over 2}
{g\alpha\over{\Tb\rhob\cp}}\nabla\Kb\nabla\Tb_{\rm ad}\right)^{-1/2},
\eqno(12)$$ where $w_0$ is the characteristic convective velocity at
the start of the overshooting zone, which can be estimated from the
local mixing length equations.  Equation (12) is essentially Zahn's
equation (3.9), except for his inclusion of some dimensionless
constants of order unity to account for the relation between $w\theta$
and $\wtb$.

Zahn's reasoning is quite different from our preceding analysis, which
presumed the viscosity term, and not the buoyancy term, dominates the
overshooting calculation.  Although many authors have made the same
presumption in the past as Zahn, Umezu (1992) demonstrates that the
viscosity term is important for the computation of overshooting.
Furthermore, the solutions in Figure 1 verify that although the
buoyancy term in equation (11) becomes negative and contributes to the
deceleration of velocities, it never overwhelms the viscosity term.
The discrepancy between Zahn's analysis and ours comes from his having
a nearly adiabatic overshooting zone compared to our nearly radiative
overshooting zone.  Ultimately, we think this can be traced to his
neglect of the term $\Kb\nabla^2\Tb$ in the expansion of
$\nabla(\Kb\nabla\Tb)$.  Indeed this term is dominant over the
$\nabla\Kb\nabla\Tb$ term in the overshooting regions of our models
below.  Presumably this term is responsible for making the temperature
gradient take nearly the radiative value, rather than the adiabatic
value, in the overshooting region.

We compare predictions for a solar model based on Zahn's and our
analyses to data on the Sun's overshooting (really undershooting)
zone.  Convection in the solar convection zone is efficient, and the
temperature gradient is nearly adiabatic.  Using standard notation for
the dimensionless logarithmic temperature gradient, we have
$(\nabla-\nabla_{\rm ad})_1\approx 10^{-6}$ in the convective zone of
the Sun (Chaboyer, private communication, for a solar model computed
using local mixing length theory with $\ell\approx 2\hp$).  Below the
convection zone, the temperature gradient takes on the radiative
value, making a transition to $(\nabla-\nabla_{\rm ad})_2\approx -0.2$
over about $0.8\hp$.  Substituting these numbers into equation (6), we
find that the convective flux decays on a scale $H_{w\theta}\sim
2\times 10^{-3}\hp$.  Since this is so much smaller than the $0.8\hp$
transition region, the temperature gradient here should depart very
little from the radiative gradient.  This is consistent with recent
helioseismological observations that show the extension of the
adiabatic region below the solar convection zone is $<0.1\hp$ and
consistent with zero (Basu, Antia, \& Narasimha 1994; Monteiro,
Christensen-Dalsgaard, \& Thompson 1994), and is in accord with the
solar model computed using Xiong's nonlocal theory (Xiong \& Chen
1992).  In contrast, Zahn predicts an extension to the region of
nearly adiabatic convection of about half a pressure scale height
below the stability transition, $H_{w\theta}\sim 0.5\hp$ (by Zahn's
eq. 3.13).

Below the very narrow region of width $H_{w\theta}$ across which the
temperature gradient becomes radiative rather than adiabatic, we
predict a decay of the turbulent velocity on a scale of
$H_{w^2}\sim\hp$, comparable to the $e$-folding scale of $0.6\hp$
found by Xiong and Chen (1992).  The extent which turbulent velocities
overshoot the stability boundary bears on the amount of lithium and
beryllium depletion in the solar atmosphere, but the precise extent of
overshooting is not probed by helioseismology.  Zahn predicts a decay
on the scale $H_{w^2}\sim H_{w\theta}\sim 0.5\hp$, to within factors
of order unity.  Although comparable with the our estimate, the
physics leading to this estimate is quite different.  In fact, since
Zahn assumes the thermal conductivity is linear at the stability
transition and we make it discontinuous, Zahn's analysis cannot
address our simple model.

\section{The Numerical Method of Solution}

We use a relaxation method to solve the nonlinear, coupled
differential equations (Press \etal 1986).  It has been necessary to
treat the temporal integration and spatial gradients in nontrivial
ways to obtain numerical solutions to the moment equations.  Although
our main interest is in the steady state solutions of the equations,
solutions usually cannot be obtained from crude initial guesses.
However, following the time evolution from an initial guess to the
steady state works well.

Convective time scales are orders of magnitude longer than
hydrodynamic time scales.  Thus, relaxing a fluid to a convective
steady state over many mixing times would be prohibitive if the time
step of the computation were Courant-limited.  To avoid such
limitations on the time step, we treat the time integration implicitly
(\cf Press \etal 1986).  That is, if the $N$ time dependent moments
are represented by $f^n_i$, where the index $i$ refers to a particular
moment and $n$ to the time step, the moments are integrated using
$$f_i^{n+1}=f_i^n+y_i(z,f_1^{n+1},...,f_N^{n+1})\Delta t.\eqno(13)$$
The source terms $y_i$ are, in general, functions of all the variables
and their spatial derivatives.  The integration is implicit because
the updated values $f_i^{n+1}$ are used in the source term.  For
general nonlinear differential equations, implicit integrations are
not guaranteed to be stable for arbitrarily large time steps.  Our
equations, however, seem stable for arbitrarily large time steps if
the solutions are sufficiently near the steady state, but not
necessarily far from the steady state.

Stability of the integration does not imply accuracy, and to maintain
reasonable accuracy, we use a time step such that no variable changes
by more than 100\% in one step, $$\Delta t=
\epsilon\min(f_i^n/y_i).\eqno(14)$$ The constant $\epsilon$ usually
need not be much less than unity to have both stability and accuracy,
even when far from the steady state.  (An exception to this time step
calculation is that if the time step is limited by a point in an
overshooting zone where some variables become very small and eq. 14
ill-defined, we take a time step of $\Delta t=\Delta z/\sigma_w$,
where $\Delta z$ is the distance between grid points.)  As the
solutions approach the steady state, the time step increases.  As a
matter of standard practice, we integrate all solutions to a time of
$10^{10}$.  (For reference, a typical sound crossing time is of order
unity.)

We adopt spatial boundary conditions that are consistent with the
reflecting boundary conditions used in the GSPH simulations.  In the
GSPH simulations, particle velocities changed sign when they hit walls
of the one-dimensional box, but retained the same temperature.  In
this case, moments that are odd in $w$ are zero at the walls of the
box, whereas moments even in $w$ have gradients that are zero at
walls.  Each equation for an odd $w$ moment has one boundary condition
placed at one of the walls.  Each equation for an even moment and the
corresponding equation defining the gradient require two boundary
conditions on the gradient, one at each wall.  One consequence of
applying these boundary rules rigorously (as is necessary for
comparison with the GSPH simulations) is that we have adopted a
gravitational acceleration $g$ which goes to zero at the boundaries.
We have boundary regions of 10\% the width of the box over which the
gravity falls from unity to zero (\cf \S3.3 of Paper II).  Heating and
cooling of the fluid occurs over these same regions.

The relaxation routines of Press \etal (1986) spatially couple only
two neighboring points at a time, so that spatial derivatives must be
taken as first differences.  For a spatially centered differencing
scheme, odd variables (ones that are zero at the boundaries) would be
located at each grid point, and even variables at half grid points,
since the derivatives of even variables are odd and the derivatives of
odd variables are even.  Although the numerical routine does not allow
this (since this differencing would couple three points), we can
construct the differencing as if even variables are at the half grid
points instead of the grid points themselves.  Boundary conditions do
not present a problem since we never put boundary conditions on even
variables.  Without this trick to simulate spatial centering of the
variables and to give second order spatial accuracy, the solutions of
the moment equations can develop sawtooth oscillations that ultimately
prevent convergence.  The solutions presented here have been solved on
a grid of 50 evenly spaced points, with the two walls representing the
first and last points.

The Xiong solution discussed above solves 6 time dependent moment
equations, includes an equation of state, and three closure relations.
Seven first derivatives are defined, as is the second derivative of
$v$, mostly because they are required for boundary conditions.  Thus,
the minimum number of coupled equations is 17.  The Full solution
discussed above solves 10 time dependent moment equations, includes an
equation of state, and requires four closure relations, 10 first
derivatives, and the second derivative of $v$.  The minimum number of
equations is 26.  For convenience we have defined a few auxiliary
variables, so that we actually solve 29 equations simultaneously.

Our numerical solutions of the moment equations use the same
dimensionless units as the GSPH simulations of Paper II.  The units
are defined by setting the acceleration of gravity $g=1$, Newton's
constant $G=1$, the mean density of the fluid averaged over the entire
box $\rhob=1$, and the ratio of Boltzmann's constant to specific mass
$\kb/\mu=1$ (\cf Appendix C of Paper II).  In these units, the
equation of state is $P=\rhob\Tb$.

\section{Comparisons Between Theory and Simulations}

Convection models have a heat source in the lower boundary region and
a heat sink for cooling in the upper boundary region.  The cooling
rate is adjusted so that the temperature of the fluid is $\Tb\approx
1$ at the bottom.  There is about one pressure scale height across the
box.  If the thermal conductivity $\Kb$ is small enough, the fluid
will convect.  If $\Kb$ is within a factor of a few of the value for
critical stability, convection is inefficient since radiative
diffusion continues to carry a significant fraction of the energy
flux.  If $\Kb$ is much smaller, convection is efficient since
convection carries nearly the entire energy flux.

Below we present solutions for two models with constant $\Kb$ across
the box, one in the efficient regime and the other in the inefficient
regime.  We call these ``homogeneous models'' due to the constant
$\Kb$, and there are no significant gradients of convective properties
on scales less than a pressure scale height.  By increasing $\Kb$ from
a small value that causes instability to a larger value that gives
stability, we construct overshooting models.  We present two
overshooting models, where the unstable regions correspond to the
efficient and inefficient regimes.  The stability transition is at the
center of the box, with the overshoot region in the upper half.  The
details of the particular parameters for each of these models can be
found in \S\S5.1--5.4 of Paper II.

\subsection{``Homogeneous'' Models}

In Figure 2 we compare the superadiabatic gradient $\dgt$ and the
three second moments (actually the velocity and temperature
dispersions $\sigma_w$ and $\sigma_\theta$, and the $\wtb$ correlation
$\wtb_n=\wtb/\sigma_w\sigma_\theta$) of the two homogeneous GSPH
simulations of Paper II with the Xiong and Full solutions.  Figure 3
compares the four third moments, and Figures 4 and 5 compare the four
fourth moments, unnormalized and normalized, required in the Full
solution.

The Xiong third moments, derived from closure equations (2) and shown
in Figure 3, are mostly in rough qualitative agreement with the GSPH
simulations in the interior of the box.  $\wwwb$ has the right
qualitative shape in both cases, but about half the GSPH value since
we use $c=1$ instead of the optimum 1.8.  The boundary regions of the
remaining third moments are grossly in error in the efficient model,
and this is reflected in $\dgt$ of Figure 2 as well.  Despite these
discrepancies, the second moments of the efficient model, including
the boundary regions, are in good qualitative and quantitative
agreement with the GSPH simulations.  The second moments hardly change
if we instead use the optimum $c's$ in the closure relations.  The
third moments of the inefficient model also show significant
quantitative discrepancies, although qualitative trends are predicted
well.  Nevertheless, the quantitative agreement of the second moments
is much better.

We consider now the Full solution.  When the $\wtb$ correlation is
unity, as is seen to be the case in Figure 2 (excepting the boundary
regions), the quasi-normal closure predicts normalized fourth moments
equal to 3\footnote{$^1$}{Normalization means dividing by powers of
$\sigma_w$ and $\sigma_\theta$ to get nondimensional ratios.  For a
pure quasi-gaussian closure with $\xi=3$, $d=0$ in eqs. (3),
$\wwwwb_n=3$, $\wwwtb_n=3\wtb_n$, $\wwttb_n=1+2\wtb_n$, and
$\wtttb_n=3\wtb_n$.  Thus, most fourth moments are not exactly 3 if
the $\wtb_n$ correlation is not unity.}.  As seen in Figure 5, the
GSPH simulations show that the normalized fourth moments actually vary
from about 4 at the bottom of the box to about 2 at the top (boundary
regions excepted, see also Fig. 15 of Paper II), and, in fact,
including the diffusion terms in equations (3) does precisely this.
The unnormalized fourth moments of the Full solution in Figure 4 are
in reasonable quantitative agreement with the GSPH results.

In Figure 3 the third moments from the Full solution more nearly
correspond to the GSPH results, both quantitatively and qualitatively,
than the Xiong model.  Oddly, the second moments are not also
superior, and, in fact, are generally inferior in the boundary
regions.  The diffusion term in closure equations (3) have had the
effect of suppressing the amplitudes of third moments, improving
agreement in Figure 3.  The smaller third moments, however, make the
second moments of Figure 2 more local in character, making agreement
worse.

Without the diffusion term in the closure relations, the amplitudes of
the third moments become several times larger, making the gradients of
third moments correspondingly larger.  These gradients impact the
second moments by causing depressions adjacent to the boundary
regions, especially at the right boundary.  In the efficient model,
the depression has a depth of about 25\%.  In the inefficient model,
the depression grows to 100\% and $\wwb$ tries to become negative.
(Note that hints of such depressions can be seen in $\sigma_w$.)  This
is clearly unphysical, and we cannot obtain steady state solutions for
the efficient model except by controlling the amplitudes of third
moments using the diffusion term.

It appears that modeling the fourth moment closures well does not mean
the second moments will be in good agreement with the simulations.  It
seems that the relations among the second, third, and fourth moments
predicted by the moment equations is not the same as the relations
predicted by the GSPH simulations.  Indeed, the second moments agree
best when the third moments do not agree well.

\subsection{Overshooting Models}

The overshooting models provide better tests of nonlocal behavior
since convective properties change on a scale shorter than a pressure
scale height.  The second, third, and fourth moments for our efficient
and inefficient overshooting models are shown in Figures 6--9.

We consider first the Xiong solutions.  In Figure 7 we see that the
Xiong closure captures most qualitative features of the GSPH
simulations.  In particular, the peak values of the third moments
occur at about the right place in the efficient model, although the
peaks are systematically displaced to the right in the inefficient
model.  In all cases, however, with the exception of $\wwwb$ in the
efficient model, the quantitative agreement is rather poor, especially
in the lower boundary region.  The agreement of the second moments is
better, as was the case with the homogeneous models.  Of greatest
physical relevance is the behavior of $\sigma_w$ and $\wtb$ in the
overshoot region.  The extension of mixing into the overshooting zone
and the reversal of the convective flux are predicted well in both
cases.  Note, however, that the superadiabatic gradient in the
convective half of the fluid is not predicted well due to problems in
the boundary region.  Both the GSPH and Xiong solutions have
$\sigma_\theta$ decreasing faster in the overshooting zone than
$\sigma_w$, as predicted by equations (6), (7), and (9).

We now consider the Full solutions.  In Paper II we showed that the
quasi-normal approximation represents the fourth moments reasonably
well in the unstable region of the overshooting models, but in
overshooting zones, fourth moments are more complicated as seen in
Figure 9 (see also Fig. 19 of Paper II).  Near the stability
transition, the normalized fourth moments grow as the $w$ and $\theta$
distributions become skewed by nonlocal effects.  Moments even in
$\theta$ grow and then decay in the overshooting zone.  Moments odd in
$\theta$ change sign in the overshooting zone, after about $0.6\ell$
for the efficient overshooting model and almost immediately for the
inefficient overshooting model.  It is remarkable that including the
diffusion term in equations (3) reproduces many features in Figure 9
(although many details are in quantitative error), even though the
diffusion term was justified as a numerical necessity, without a
physical basis.  In Figure 8, since amplitudes are small in the
overshooting zone, the appearance of errors in the overshooting zone
is suppressed, although for the inefficient overshooting model, errors
in the unstable region are large.

In the efficient overshooting model the third moment solutions in
Figure 7 generally agree better with the GSPH results than the Xiong
solution.  As with the homogeneous models, the suppressed third moment
amplitudes make the efficient second moments in Figure 6 more local in
character.  Indeed, the second moments of the Xiong solution seem
superior to those of the Full solution.  It appears that better
agreement between the Full and GSPH second moments could be achieved
by reducing the magnitude of the diffusion terms in equations (3).  Of
course, the third moments would then be worse.

For the inefficient overshooting model, the third moments of the Full
solution are systematically too big.  The second moments, however, are
in reasonable agreement, and the scale of overshooting predicted by
the slope of $\sigma_w$ is approximately correct.  Increasing the
magnitude of the diffusion terms in equations (3) could improve
agreement of the third moments, but would degrade the second moments.

If we solve the equations using a pure quasi-normal closure, that is,
without the diffusion of third moments, the resulting third moment
curves have qualitatively similar shapes, but roughly twice the
amplitude.  The larger third moments gradients try to make $\wwb$ and
$\ttb$ negative near the left boundary, causing the numerical scheme
to fail.  Without the diffusion term, we cannot obtain steady-state
solutions for either the efficient or inefficient overshooting models.

Modeling the fourth moments reasonably well, as in the efficient
overshooting model, has not caused the third moments to agree with the
GSPH results to a comparable degree.  Likewise, modeling the fourth
moments very badly, as in the inefficient overshooting model, has not
caused the third moments to be comparably bad.  Even though the
detailed agreement of the third moments is not great for the Xiong or
Full solutions, the agreement of the second moments is considerable
better.  As with the homogeneous models above, the relations among the
GSPH moments are not the same as the relations predicted by the Xiong
or Full solutions.

\subsection{Comparison with Ballistic Trajectories}

Theoretical investigators of overshooting have most often performed a
ballistic particle sort of analysis, where the equations of motion are
integrated until the velocity of an overshooting particle is zero
(Shavis \& Salpeter 1973 is a classic example).  We do the same here
to highlight a fundamental difference between this and the moment
equation approach.  The equations of motion are
$$\eqalignno{\dot z=&w&(15a)\cr
\dot w=&g\alpha\theta/\Tb-(A+B\vert w\vert)w,&(15b)\cr
\dot\theta=&\dgt w-(D+E\vert w\vert)\theta.&(15c)\cr}$$
We have replaced the turbulent damping rates $B\sigma_w$ and
$E\sigma_w$ with $B\vert w\vert$ and $E\vert w\vert$, since $w$
represents a characteristic turbulent velocity in the ballistic
calculation.  Furthermore, there is no stable attractor in a
convective region unless the equations are made nonlinear in this way.
The constant $\alpha$ is the coefficient of thermal expansion defined
in Appendix A.  We use the background values of $\Tb$ and $\dgt$
computed for the Full solutions, and use as initial values of $w$ and
$\theta$ the local solutions of $\sigma_w$ and $\sigma_\theta$.

Results of these integrations for both the efficient and inefficient
overshooting models are shown in Figure 10.  In each case we show two
integrations, one starting at $z=0.4$ as the solid curves and another
starting at $z=0.3$ as the dotted curves.  In the top panels we show
the evolution in $w$-$\theta$ phase space.  The particles enter the
overshoot region with positive $w$ and $\theta$.  In the case of the
efficient overshooting model, the particles spiral toward the origin
of phase space at the Brunt-V\"ais\"al\"a frequency,
$\omega=(g\alpha\vert\dgt\vert/\Tb)^{1/2}$, with an amplitude decaying
at the turbulent damping rate.  In the inefficient overshooting model,
damping by radiative diffusion is faster than turbulent damping (\ie,
$D>E\sigma_w$), with a rate comparable to the Brunt-V\"ais\"al\"a
frequency, so that particles evolve directly to the origin.

The remaining panels in Figure 10 show $w$ versus $z$ and $\theta$
versus $z$.  In ballistic calculations of overshooting (\eg, Shaviv \&
Salpeter 1973; Maeder 1975; Schmitt, Rosner, \& Bohn; Langer 1986;
Zahn 1991), the overshooting distance usually is defined by the
location where $w$ first reaches zero.  If we take the stability
transition at $z=0.35$ for the efficient overshooting model and
$z=0.4$ for the inefficient overshooting model, as suggested by the
nonlocal $\dgt$ curves, overshooting distances are $\dover\approx
0.6\ell$ in both cases.  Reversal of the convective flux is predicted
where $\theta$ becomes negative, after about $0.3\ell$ and $0.05\ell$
in the efficient and inefficient cases.

The scale for $w$ to reach zero is comparable to the scale for
$\sigma_w$ to decay by a factor $e$.  The nonlocal solutions for
$\sigma_w$, however, approach zero asymptotically, but never reach it
within a finite distance, as the ballistic calculations do.  Although
an $e$-folding distance might represent a realistic scale for the
decay of turbulent velocities, since convective fluids generally mix
so much faster than nuclear time scales, overshooting may be important
for many $e$-folding distances.  To define overshooting distances, it
will be necessary to compare the mixing time scales with nuclear
evolution time scales.  Defining a hard edge to overshooting where
$w=0$ is probably misleading, unless overshooting is negligible and
convection is essentially local.

\section{Equations Modified for Scattering}

Rather than deal with turbulent losses using eddy viscosity and eddy
diffusion rates in the dynamical equations for $w$ and $\theta$, we
can treat turbulent losses as a scattering processes in our Boltzmann
transport formulation of convection.  In addition, although we have
neglected previously turbulence in the horizontal dimensions, we
include it now.  Thus, we consider the evolution of particles in a
$w$-$u$-$\theta$ phase space, where $u$ is a horizontal velocity
perturbation.  The velocity dispersion is now defined as
$\sigma=(\wwb+2\uub)^{1/2}$, and the ensemble of particles is
described by the distribution function $\fr(t,z,w,u,\theta)$.

The alternate treatment of turbulent losses in the collision term
allows for greater flexibility through the particular choices we make
in modeling the scattering process.  Here we assume that scattering is
isotropic in velocity, so that a single parameter $\xi$ defines the
width of a gaussian scattering function for both $w$ and $u$.  A
parameter $\varpi$ defines the width of a gaussian scattering function
for $\theta$.  In Paper I we carried out this alternate derivation to
the local level only.  This approach was used by Narayan, Loeb, and
Kumar (1994) to study causal diffusion and by Kumar, Narayan, and Loeb
(1995) to study the interaction of convection and rotation. In
Appendix B of this paper, we write the complete moment equations, up
to third order.  We see that in the appropriate limit, the second
moment equations are identical to the previous formulation, but the
third and higher moment equations are necessarily different.

We solve these alternate equations using the same closure schemes as
above, but supplemented with a few more closure relations for new
moments of the horizontal velocity perturbation $u$.  We shall refer
to these alternate equations as the scattering equations, and to our
two closure schemes as the Xiong and Full solutions, as before.

\noindent{\it Xiong Solution}--If we close the equations at the
level of the third moments using the diffusion approximation, in
addition to equations (2) we require another closure,
$$\wuub=-c_4\ell\sigma_w\nabla\uub.\eqno(16)$$ We take the constant
$c_4=1$, like the other constants in equation (2).

\noindent{\it Full Solution}--The Full solution requires additional
fourth moment closures.  Using the same modified form of the
quasi-normal approximation as in equations (3), we have
$$\eqalignno{
\wwuub&=\zeta\wwb\,\uub/3-d\ell\sigma_w\nabla\wuub&(17a)\cr
\wuutb&=\zeta\wtb\,\uub/3-d\ell\sigma_w\nabla\uutb&(17b)\cr
\uuuub&=\zeta\uub^2&(17c)\cr
\uuttb&=\zeta\uub\,\ttb/3.&(17d)\cr}$$
In writing these relations we have used $\utb=\wub=0$, since there is
no preferred horizontal direction.  We include diffusion terms on
equations (17a) and (17b) since these fourth moments are associated
with the turbulent transport term of third moment equations.  The
closures of equations (17c) and (17d) are required for certain terms
not directly associated to a particular third moment.  As before, we
use $\zeta=3$ for quasi-normal closure, and $d=0.8$.

The Xiong solution involves 7 time dependent moment equations,
requires an equation of state, 4 closure relations, and 10 derivatives
for boundary conditions, making the minimum problem one of 22 coupled
equations.  The Full solution uses 13 time dependent moment equations,
an equation of state, 8 closures, and 18 derivatives for boundary
conditions, making 40 equations in all.  Because we have added a few
auxiliary variables for convenience, we actually solve 43 equations
simultaneously.

\section{Comparisons Between the Scattering Theory and Simulations}

We present results for models using precisely the same four sets of
parameters as above.  We also present results of simulations from a
version of the GSPH code modified to be consistent with this alternate
formulation of the moment theory.  The new equations and the
corresponding modifications to the GSPH code are discussed in detail
in Appendix B.  We mention the main modifications to the GSPH code
here.  The evolutionary equations for particles no longer have the
eddy damping terms, and particles have a probability of scattering
with a characteristic time scale $\ell/\sigma$.  If a particle does
scatter, it does so randomly into distributions of $w$, $u$, and
$\theta$ described by parameters $\xi$ and $\varpi$.

To understand the physical implications for various $\xi$, $\varpi$,
we performed GSPH simulations for a range of these parameters.  If
$\xi=\varpi=0$, there are large peaks at the origin of the phase space
distribution, since all particles eventually are scattered there.
Then any convective flux must be carried by broad tails of the
distribution function $\fr(z,w,u,\theta)$.  In this case, normalized
fourth moments are of order 50, instead of closer to the quasi-normal
value of 3.  Not only do the GSPH results seem unreasonable if
$\xi=\varpi=0$, but we cannot obtain steady-state solutions to the
moment equations in this case.  Certain variables which must be
positive tend to become negative as the moment equations evolve toward
the steady state.

If we use $\xi=\varpi=0.9$, the normalized fourth moments generated by
GSPH simulations are more reasonable, having values of order 10 or
less in convective regions.  Furthermore, we can obtain steady state
solutions of the moment equations for all four of our model problems.
Hence, we adopt $\xi=\varpi=0.9$ in the following calculations.  Below
we compare these solutions to results of new GSPH simulations.  We
also consider whether these comparisons are more or less favorable
than the comparisons of \S4.

\subsection{``Homogeneous'' Models}

The second moments of the homogeneous models are compared with the new
GSPH results in Figure 11.  At the cost of some confusion, we have
plotted both the vertical and horizontal velocity dispersions,
$\sigma_w$ and $\sigma_u$, in the same panels.  The curves for
$\sigma_u$ are always the lower set.  The third moments are shown in
Figure 12.  The moments $\wuub$ and $\uutb$ are not shown to avoid
overwhelming confusion.  They are always much smaller than third
moments with $u$ replaced by $w$.  The fourth moments are shown in
Figures 13 and 14, but here too we have not plotted moments involving
$u$.

The GSPH third moments in Figure 12 are qualitatively similar to those
of Figure 3.  In the efficient model, the Xiong third moments show
significant qualitative failures, particularly associated with the
boundary regions.  The boundary problem is seen also in the $\dgt$
curve in Figure 11.  Despite these differences, the second moments,
including the boundary regions, are in good qualitative agreement with
the simulations.  The most noteworthy quantitative difference is that
the velocity dispersions $\sigma_w$ and $\sigma_u$ are predicted to be
$\sim 25\%$ to big.  We note that the velocity-temperature correlation
is less than unity, as predicted by equation (B25) which gives
$\wtb/\sigma_w\sigma_\theta=0.55$ in the local limit.  In the
inefficient model, the third moments are in excellent agreement with
the simulations and are clearly better than those of Figure 3.
Nevertheless, the agreement of the second moments is not improved by a
corresponding degree.  Again the velocity dispersions are
systematically too big, and $\sigma_\theta$ does not have quite the
right shape.  We note also that although equation (B26) predicts that
$\wtb/\sigma_w\sigma_\theta=1$ in the inefficient regime, the computed
value is reduced somewhat, because the large value of $\xi$ has moved
this inefficient model closer to the efficient regime (see eq. B27).

We now consider the Full solutions.  In Figures 13 and 14, the
predicted fourth moments are generally smaller than those of the GSPH
simulations.  The quantitative agreement is, however, worse than in
the corresponding Figures 4 and 5.  In Figure 12, certain third
moments are improved over the Xiong solution, particularly in the
efficient model.  The quality of third moments are comparable to those
in corresponding Figure 3.  As usual, the second moments in Figure 11
exhibit too much local character.

We find that third moments can be much improved if $d=0.2$ in the
closure relations instead of $d=0.8$, restoring some of the
nonlocality to the second moments.  With this change, the fourth
moments hardly change at all.  Apparently small and subtle changes in
the fourth moments can have much more dramatic effects on second and
third moments.  Big changes in the fourth moments, making them appear
in much better agreement, would lead to much worse solutions of the
lower moments.

\subsection{Overshooting Models}

Solutions of the scattering theory with $\xi=\varpi=0.9$ for the
overshooting models are shown in Figures 15--18, where we present also
the second, third, and fourth moments of new GSPH simulations.

We discuss first the Xiong solutions.  The Xiong third moments in
Figure 16 capture the behavior of the GSPH simulations only roughly.
As we observed with the original version of the theory in \S4,
however, the agreement of the second moments is better.  Nevertheless,
we do note a few important discrepancies.  The velocity dispersions
$\sigma_w$ and $\sigma_u$ are somewhat to big in the convective half
of the fluid, and $\sigma_w$ decays too fast in the overshooting zone.
In the efficient overshooting model, the superadiabatic gradient is in
error in the left boundary region.  In the inefficient overshooting
model, the shape of the $\wtb_n$ correlation is not captured well,
which can be attributed to having $\sigma_\theta$ too small in the
overshooting region.

In both overshooting models, $\sigma_w$ decays faster in the
overshooting region than does $\sigma_u$.  The reason is that negative
buoyancy and turbulent viscosity decelerate vertical motions, while
only turbulent viscosity acts horizontally.  Thus, the ratio of
kinetic energy in the horizontal motions to kinetic energy in vertical
motions, $\uub/\wwb$, increases in the overshooting zone.

The fourth moments of the Full solution in Figures 17 and 18 are not
in good agreement with the GSPH simulations.  In Figure 18, however,
the normalized moments do show many qualitatively correct features.
In the efficient overshooting model, the difference with $\wwwwb$ is
particularly severe.\footnote{$^2$}{We note that the GSPH $\wwwwb$
curve hardly decays in the overshooting region of the efficient
overshooting model.  In the overshooting region, $\sigma$ decreases,
thus increasing the characteristic time scale for scattering.
Overshooting particles with the largest $w$ and $\theta$ can cross the
overshooting zone without scattering, accelerating for most of the
distance, whereas more typical particles do scatter to smaller
amplitudes.  Although $\sigma_w$ is decreasing, the distribution of
$w$ has an envelope that actually gets broader in the overshooting
zone, indicative of broad wings in the distribution of $w$ that
maintains the magnitude of $\wwwwb$ even as $\wwb$ decreases.}  The
solutions for the inefficient overshooting model are worse.  Despite
these large discrepancies of fourth moments, in Figure 16 the third
moments of the Full solution are in better agreement, although not
better than corresponding Figure 7.  In the efficient overshooting
model, they are clearly superior to the Xiong third moments, and in
the inefficient overshooting model, agreement is comparable (the
amplitudes are slightly worse, but the location of the peaks of the
curves is better).  As usual, the second moments in Figure 15 have too
much local character, as evidenced especially in the efficient
overshooting panels.  This suggests that decreasing the magnitude of
the diffusion term on the fourth moment closures could improve
agreement at the second moment level.  Indeed, if we use $d=0.2$
instead of $d=0.8$, we find $\sigma_w$ and $\sigma_\theta$ in better
agreement in the inefficient overshooting model (although the
qualitative shape of the normalized $\wtb$ curve and the amplitudes of
the third moments are slightly worse), but we cannot get a converged
solution to the efficient overshooting model.

We can describe third moments better using the Full solution, rather
than the Xiong closures.  As before, however, good third moments do
not necessarily imply good second moments.  Indeed, there are
substantial differences between the Xiong and Full third moments, but
the differences between the Xiong and Full second moments are not as
great.

\section{Discussion and Conclusions}

We have solved the moment equations describing stellar convection for
two, somewhat different, formulations of nonlocal mixing length
theory.  In the first formulation, velocity and temperature
perturbations were damped at an eddy diffusion rate defined by the
mixing length.  The second formulation treated turbulent dissipation
as a scattering process, where the scattering time scale is determined
by the time required to cross the distance of a mixing length.  In
both cases we solved the equations using two different closure
schemes, called the Xiong and Full solutions.  The Xiong solutions
closed the equations at the third moments using the diffusion
approximation.  The Full solutions close the equations at the fourth
moments with a modified version of the quasi-normal approximation.
These various solutions were computed for four models: efficient and
inefficient homogeneous convection, and efficient and inefficient
overshooting.

Regardless of the closure scheme, certain features of our nonlocal
solutions seem to hold generally.  In an overshooting zone, shortly
beyond the stability boundary, the convective flux becomes negative.
Hence, more energy must be transported outward by radiation than in
local convection, and in the overshooting region where $\dgt<0$, the
temperature gradient becomes somewhat steeper and more nearly
adiabatic.  In our solutions, the convective flux does not obtain
large negative values in the overshooting zone, and the departure of
the temperature gradient from the local radiative value is small.
Thus, overshooting hardly affects the hydrostatic structure of a star.
This conclusion is supported by recent helioseismological results, but
is contrary to the claim of many authors that the overshooting zone is
nearly adiabatic.  If the overshooting zone were nearly adiabatic,
overshooting blobs moving against a barely-stable temperature gradient
could not obtain large temperature deficits that give a large negative
flux, as would be required for big departures from the radiative
temperature gradient.

In the overshooting zone, the turbulent velocities do not decay as
fast as the convective flux.  Solutions of the Paper I equations have
a turbulent velocity $\sigma_w$ that decays by a factor $e$ in a
distance $0.5\ell$--$0.8\ell$, depending on the model.  The scattering
equations give a decay scale about twice as large.  Turbulent
viscosity is important in damping turbulent velocities; the
deceleration of negative buoyancy is less important.  This is
different from most previous overshooting calculations, which regard
negative buoyancy as the dominant or only mechanism for deceleration.
These results about the convective flux and large extent of
overshooting are qualitatively consistent with those of Xiong (1985)
and Xiong \& Chen (1992).

Although the moment equations predict a characteristic distance for
the decay of turbulent velocity, the turbulent velocity approaches
zero asymptotically and the overshooting distance is formally
infinite.  This is fundamentally different from ballistic particle
theories that compute an unambiguous, finite overshooting distance.
This difference is a consequence of considering the entire ensemble of
turbulent blobs simultaneously, rather than evolving only one blob in
a static background.  The ballistic calculation does seem useful for
defining characteristic scales, although the definite boundaries on
the extent of overshooting may be misleading.  It is possible that
turbulent mixing may be more rapid than relevant nuclear burning time
scales for several scale distances, so that overshooting distances
cannot be defined without considering evolutionary time scales for
detailed stellar models.  Such considerations go beyond the scope of
the present work.

In addition to simply solving the equations, we wanted to demonstrate
the internal consistency of our theory by showing that the solutions
of the moment equations reproduce the results of GSPH simulations.
Since the GSPH code and moment equations rely on the same
approximations and same picture of mixing length convection, we
anticipated that if the closure relations for high order moments were
good enough, then the solutions of the lower moments automatically
would agree with the GSPH moments.  We have demonstrated broad
qualitative agreement, and the quantitative solutions are perhaps
reasonable zeroth order approximations.  Nevertheless, detailed
comparison with the GSPH results reveals many shortcomings.  In
particular, fitting high order moments well apparently does not mean
that the lower order moments will exhibit agreement of comparable
quality.  Conversely, low order moments may agree well when high order
moments are badly in error.  As a general rule, the Xiong solution
predicts second moments better, with third moment agreement not as
good.  The Full solution predicts third moments better, but the second
moments show inferior agreement.  Furthermore, there is no strong
reason to prefer the solutions of the Paper I equations solved in \S4
or solutions of the scattering equations solved in \S6.  Since the
second moments $\wwb$ and $\wtb$ are most important for constructing
stellar models, we conclude that the Xiong closures perform
impressively well.

It is surprising to us that the quality of internal consistency is not
much better.  Whatever the degree of internal consistency, however,
the mixing length approximations are severe, and an external
consistency with the real world is not guaranteed.  The GSPH and
moment solutions eventually should be compared to and calibrated by
three-dimensional hydrodynamic simulations.  It is now well-known that
compressible convection exhibits qualitative features that the moment
theory cannot describe.  In particular, convective flows are
characterized by deeply penetrating plumes of material, with broader,
gentler upflows (Cattaneo \etal 1991; Hossain \& Mullan 1991; Hurlburt
\etal 1986; Stein \& Nordlund 1989).  Although this horizontal
structure is beyond our ability to predict, we do predict the velocity
asymmetries that lead to a downward-directed kinetic energy flux in
convective regions.  Simulations of overshooting (Singh \etal 1994)
and undershooting (Hurlburt \etal 1994) are less common, but it is
reasonable to expect that fewer and fewer plumes penetrate to
increasing overshoot distances.  The decreasing rate of turbulent
mixing at deeper overshoot distances may be described by our moment
theory.

In summary, we have a theory of nonlocal convection with enough
qualitative success that it may represent a significant improvement
over previous theories, but it clearly has significant shortcomings
too.  The quantitative disagreement with the GSPH results may have two
different origins.  1) The closure approximations may not adequately
describe the relations among high and low order moments.  2) The
moment equations themselves may be flawed.  We consider these two
possibilities in turn.

The closure approximations we use, for both the Xiong solution and
Full solution, are reasonable approximations to the actual high order
moments, as demonstrated in Paper II, but they are not the best
relations suggested by the GSPH simulations.  Our preferred third
moment closures from Paper II are an alternate form of the diffusion
approximation.  Those alternate closures, however, do not have the
required boundary symmetries, and are not compatible with the method
of solution in this paper.  The best fourth moment closures from Paper
II related fourth moments to second moments using the quasi-normal
approximation, with perturbations by third moments.  We have not been
able to solve any problems with these optimal closures, and have not
been able to solve some with the much simpler pure quasi-normal
approximation.  Failure to obtain solutions usually means that
physically positive second moments, $\wwb$ or $\ttb$, become negative
and the numerical method eventually fails to converge.  (We note that
there is nothing in the mathematical nature of the moment equations
that enforces $\wwb$ or $\ttb$ to be positive, or the correlation
$\vert\wtb/\sigma_w\sigma_\theta\vert$ to be less than unity.)

We found that adding a diffusion term to the quasi-normal closure
prevents time evolution to unphysical solutions.  Regrettably, such a
diffusion term was not investigated in Paper II, and we can think of
no physical argument justifying it, except that it seems required to
obtain numerical solutions.  Nevertheless, the third moment diffusion
term seems to introduce qualitatively desirable features into the
fourth moments, giving better results than a pure quasi-gaussian
closure.  We speculate that just as each Xiong closure contains only
one out of many terms of the complete third moment closure (given by
solution of the third moment equations), yet gives reasonable results,
so too does the third moment diffusion term, being only one of many
(given by solution of the fourth moment equations) give reasonable
results.  The third moment diffusion term has the effect of
constraining the amplitude of the third moments, thereby suppressing
nonlocal effects and giving physical solutions.  Since we cannot
obtain solutions using closures that we claimed in Paper II were very
good, we do not attribute the discrepancies between the GSPH results
and the solutions directly to the closure relations.

The other possibility, that there is some problem with the moment
equations, seems very possible.  In this paper, we presented solutions
for two different formulations of the moment theory.  Although the two
sets of equations are very similar (mainly only certain coefficients
are different), there are significant quantitative differences between
the two sets of solutions.  We have no physical basis for preferring
one formulation over the other, which suggests that the coefficients
of the various terms of the moment equations may be subject to
eventual refinement.  Finally, when the equations fail to converge,
the problem usually originates in or adjacent to the boundary regions.
This may indicate that the main problem with the equations is in the
odd choice of boundary conditions, which were adopted because no other
set of self-consistent boundary conditions was apparent for the GSPH
code.

Finally, we have not yet computed models that describe realistic
stars.  The key modification will be the use of realistic opacities in
the computation of a self-consistent thermal conductivity $\Kb$.  It
is reasonable to expect, however, that the qualitative features
discussed here, namely a nearly radiative overshooting zone of
significant extent, will remain true in more detailed models.
Although we are in agreement with Xiong (1985) and Xiong \& Chen
(1992) on these points, we are in conflict with the conclusions of
Stothers (1991) and Stothers \& Chin (1992), who argue that
overshooting is not required to understand various features of stars
and star clusters.  We have not included the effects of rotation
(cf. Kumar \etal 1995) or magnetic fields in the physics of our moment
equations, and if Stothers' conclusions prove true, then these
simplifications may explain the discrepancy.  At this time, however,
the empirically determined extent of overshooting remains
controversial.  The results of this series of papers provides a way to
make predictions for many different contexts, and it is possible that
authors who favor large overshooting and ones who favor small
overshooting may both be right if they are considering convection in
different contexts.

\ack{Acknowledgments.} I thank Ramesh Narayan for his participation
in this long-term project.  With respect to this paper, certain key
ideas about the numerical method and ideas about the scattering
formulation of the moment equations are due to him.  I am grateful to
James Chiang for offering his careful reading of and comments on the
manuscript, and to Brian Chaboyer for providing me with a
state-of-the-art solar model.

\vfill\eject
\appendix{A}{The Simplified Paper I Moment Equations}

Paper I contains a detailed derivation of the moment equations.  In
this appendix we present the slightly simplified version of them that
we solve in \S4.  Since we are concerned mainly with steady-state
solutions, we drop terms that include the mean flow velocity $\vb$
(except for the viscosity term in the momentum eq. A2), including all
advection terms.  The thermal conductivity $\Kb$ and heating rate
$\Qdb$ may be functions of vertical position $z$, but do not have any
explicit dependence on thermodynamic variables (so that we can ignore
terms with $K,_T$ and $\dot Q,_T$).  The equations are:

zeroth moment equation:
$${\partial\rhob\over\partt}+\popz(\rhob\,\vb)=0.\eqno(A1)$$

$v$ moment equation: $${\partial\vb\over{\partial
t}}+g+{1\over\rhob}\popz(P+\rhob\wwb) -C{\partial^2\vb\over{\partial
z^2}}=0.\eqno(A2)$$

$T$ moment equation: $$\eqalignno{ {\partial\Tb\over\partt}&
-{1\over{\rhob\cp}}\popz\left(\Kb{\partial\Tb\over\partz}\right)
+{1\over{\rhob\cp}}{\popz}(\cp\rhob\wtb)
-{\alpha^2\over{\rhob\cp\Tb}}\pPopz\wtb
-{1\over\cp}(A+B\sigma_w)\wwb\cr&-{\Qdb\over{\rhob\cp}}=0.&(A3)\cr}$$

$w^2$ moment equation:
$${\partial\wwb\over\partt}+{1\over\rhob}\popz(\rhob\wwwb)
+{2\alpha\over{\rhob\Tb}}\pPopz\wtb+2(A+B\sigma_w)\wwb=0.
\eqno(A4)$$

$w\theta$ moment equation: $$\eqalignno{
{\partial\wtb\over\partt}&+{1\over\rhob}\popz(\rhob\wwtb)
+{\alpha\over{\rhob\Tb}}\pPopz\ttb
+(A+B\sigma_w+D+E\sigma_w)\wtb-\dgt\wwb\cr
&-{\alpha^2\over{\rhob\cp\Tb}}\pPopz\wwtb
-{1\over\cp}(A+B\sigma_w)\wwwb=0.&(A5)\cr}$$

$\theta^2$ moment equation: $$\eqalignno{
{\partial\ttb\over\partt}&+{1\over\rhob}\popz(\rhob\wttb)-2\dgt\wtb
+2(D+E\sigma_w)\ttb-{2\alpha^2\over{\rhob\cp\Tb}}\pPopz\wttb\cr
&-{2\over\cp}(A+B\sigma_w)\wwtb=0.&(A6)\cr}$$

$w^3$ moment equation:
$${\partial\wwwb\over\partt}+{1\over\rhob}\popz(\rhob\wwwwb)
+{3\alpha\over{\rhob\Tb}}\pPopz\wwtb+3(A+B\sigma_w)\wwwb
-{3\wwb\over\rhob}\popz(\rhob\wwb)=0.\eqno(A7)$$

$w^2\theta$ moment equation: $$\eqalignno{
{\partial\wwtb\over\partt}&+{1\over\rhob}\popz(\rhob\wwwtb)
+{2\alpha\over{\rhob\Tb}}\pPopz\wttb
+[2(A+B\sigma_w)+D+E\sigma_w]\wwtb\cr
&-{2\wtb\over\rhob}\popz(\rhob\wwb)-\dgt\wwwb
-{\alpha^2\over{\rhob\cp\Tb}}\pPopz(\wwwtb-\wwb\,\wtb)\cr
&-{1\over\cp}(A+B\sigma_w)(\wwwwb-\wwb^2)-{\wwb\over\rhob}\popz(\rhob\wtb)
=0.&(A8)\cr}$$

$w\theta^2$ moment equation: $$\eqalignno{
{\partial\wttb\over\partt}&+{1\over\rhob}\popz(\rhob\wwttb)
+{\alpha\over{\rhob\Tb}}\pPopz\tttb
+[A+B\sigma_w+2(D+E\sigma_w)]\wttb-{\ttb\over\rhob}\popz(\rhob\wwb)\cr
&-2\dgt\wwtb-{2\alpha^2\over{\rhob\cp\Tb}}\pPopz
(\wwttb-\wtb^2)-{2\over\cp}(A+B\sigma_w)(\wwwtb-\wwb\,\wtb)\cr
&-{2\wtb\over\rhob}\popz(\rhob\wtb)=0.&(A9)\cr}$$

$\theta^3$ moment equation: $$\eqalignno{
{\partial\tttb\over\partt}&+{1\over\rhob}\popz(\rhob\wtttb)
-3\dgt\wttb+3(D+E\sigma_w)\tttb
-{3\alpha^2\over{\rhob\cp\Tb}}\pPopz(\wtttb-\wtb\,\ttb)\cr
&-{3\over\cp}(A+B\sigma_w)(\wwttb-\wwb\,\ttb)
-{3\ttb\over\rhob}\popz(\rhob\wtb)=0.&(A10)\cr}$$

In these equations, $g$ is the acceleration of gravity, $\cp$ is the
specific heat at constant pressure, and
$\alpha=-(\partial\ln\rho/\partial\ln T)_P$ is the coefficient of
thermal expansion.  The r.m.s. turbulent velocity is
$\sigma_w=(\wwb)^{1/2}$, and the superadiabatic gradient
$\dgt=(\alpha/\rhob\cp)\partial P/\partz -\partial\Tb/\partz$.  The
coefficients $A$, $B$, $C$, $D$, and $E$ depend on several, possibly
different length scales.  For simplicity, we make the standard mixing
length assumption that turbulent fluid blobs have the same horizontal
and vertical dimensions, $\lh$ and $\lv$, which also are equal to the
turbulent damping scales of $w$ and $\theta$, $\ell_w$ and
$\ell_\theta$.  Thus, in terms of a single mixing length $\ell$, they
are defined as $$\eqalignno{ A&=10\numic/3\ell^2,&(A11a)\cr
B&=2/\ell,&(A11b)\cr C&=4\numic/3,&(A11c)\cr
D&=3\Kb/\rhob\cp\ell^2,&(A11d)\cr E&=2/\ell.&(A11e)\cr}$$ The
coefficient $C$ only appears in the momentum equation (A2), where it
is responsible for damping background motion.  The combinations
$B\sigma_w$ and $E\sigma_w$ define the rates of turbulent (or eddy)
damping of $w$ and $\theta$, while $A$ and $D$ define their damping
rates by microscopic processes (\ie, by viscosity and radiative
diffusion).  The ratios of turbulent to microscopic diffusion rates
are defined by the Reynolds number $Re=B\sigma_w/A$ and the Peclet
number $Pe=E\sigma_w/D$.

\vfill\eject
\appendix{B}{The Alternative Scattering Equations}

In this appendix we outline an alternative derivation of the moment
equations.  In this formulation of the Boltzmann transport theory,
turbulent losses are included through the collision term of the
transport equation.  In Paper I we derived only the local equations
for this alternate formulation of the theory. Those results are
extended to the nonlocal level of the theory here.  It proves easy to
include turbulence in the horizontal as well as vertical dimensions,
so we consider a three-dimensional phase space $v_x$-$v_z$-$T$, where
$v_x$ and $v_z$ are horizontal and vertical velocities, respectively.
The distribution function $\fa(t,z,v_x,v_z,T)$ evolves according to
$${\partial\fa\over\partt}+\popz(v_z\fa)+{\partial\over{\partial
v_x}}(\dot v_x\fa)+{\partial\over{\partial v_z}}(\dot v_z\fa)
+{\partial\over{\partial T}}(\dot T\fa)=\Gamma^+-\Gamma^-.\eqno(B1)$$
The perturbations with respect to the mean background are
$w=v_z-\vb_z$, $u=v_x-\vb_x$, and $\theta=T-\Tb$.  Since no bulk
forces act horizontally, $\vb_x$=0.  The three-dimensional velocity
dispersion is $\sigma=(\wwb+2\uub)^{1/2}$.  We do not require a
horizontal spatial coordinate since there are no horizontal gradients
in our problem.

The collision term on the right-hand side of equation (B1) is divided
into destruction and creation functions.  Particles lose their
identity due to interactions with the rest of the fluid on a
characteristic time scale $\sim\ell/\sigma$ in mixing length
convection, so we model the destruction term as $$\Gamma^-=2B\sigma
\fa.\eqno(B2)$$ The constant $B$ sets the rate of collisional
scattering.  Fluid blobs are not destroyed in collisions, but
scattered to new locations in $v_x$-$v_z$-$T$ phase space.  Thus, the
creation function must have the form $$\Gamma^+=2B\sigma
f_0,\eqno(B3)$$ where $f_0$ describes the distribution of velocities
and temperatures into which fluid blobs get scattered.  The
flexibility of this formulation of the theory lies in our freedom to
model the function $f_0$.

In the derivation presented here, we parametrize the creation function
$f_0$ by its second moments according to
$$(\wwb)_0=(\uub)_0={\xi^2\sigma^2\over 3},\eqno(B4a)$$
$$(\ttb)_0=\varpi^2\theta^2,\eqno(B4b)$$
$$(\wtb)_0=(\overline{wu})_0=(\overline{u\theta})_0=0,\eqno(B4c)$$
where $\xi$ and $\varpi$ are new parameters of the theory.  Equation
(B4a) asserts that particles are scattered isotropically, with a
velocity dispersion that is some fraction of the initial velocity
dispersion.  Equation (B4b) makes the temperature dispersion of
scattered particles a fraction of the initial temperature dispersion.
The remaining second moments and all higher moments of $f_0$ are
chosen to be zero.  As long as $\xi<1$, $\varpi<1$, the scattering
term will damp turbulent velocity and temperature excesses, and not
enhance them.

Here we present mainly the results of the derivations, since a
detailed description of the mathematical formalism and underlying
philosophy can be found in Paper I.

\centerline{\sectionfont B.1. The $\dot v_z$, $\dot v_x$, $\dot T$, and First
Moment Equations}

The dynamical equations that describe the evolution of a fluid blob
are $$\eqalignno{\dot v_z=&-g-{1\over\rhob}\pPopz
-{\alpha\theta\over{\rhob\Tb}}-Aw+C{\partial^2\vb_z\over{\partz^2}},&(B5)\cr
\dot v_x=&-Au,&(B6)\cr
\dot T=&{\alpha\over\rhob\cp}{dP\over{dt}}\left(1+{\alpha\theta\over\Tb}\right)
+{1\over{\rhob\cp}}\popz\left(\Kb{\partial\Tb\over\partz}\right)
-D\theta+K,_T{\partial^2\Tb\over{\partz^2}}\theta\cr
&+{1\over\cp}(A+B\sigma)(w^2+2u^2)-{1\over\cp}B\xi^2\sigma^3
+{C\over\cp}\left(\pvopz\right)^2 +{\Qdb\over{\rhob\cp}}+{\dot
Q,_T\theta\over{\rhob\cp}}.&(B7)\cr}$$ These equations are very
similar to the dynamical equations of our original formulation, with
the significant difference that there are no turbulent damping terms.
Whereas previously vertical velocity perturbations were damped by both
microscopic and turbulent viscosity with a term of the form
$(A+B\sigma_w)w$, in equation (B5) we have only the microscopic term
$Aw$, as similarly in the horizontal velocity equation (B6).  Also,
whereas temperature perturbations were damped by microscopic
(radiative) and turbulent diffusion with a term $(D+E\sigma_w)\theta$,
in equation (B7) only the microscopic term $D\theta$ is used.
Equation (B7) also has been modified to correctly include the viscous
heating associated with the collision term.

The GSPH code, described in detail in Paper II, can accommodate this
alternate treatment of turbulent losses easily.  The one-dimensional
code described in Paper II has been expanded to include horizontal
velocities, and the dynamical equations have been modified from the
original relations to these alternate ones.  Although there was no
scattering in the original code, we have added a routine to scatter
particles.  Particles scatter on a characteristic time scale
$\tau=(2B\sigma)^{-1}$, so the probability for a particle to scatter
in a time step $\Delta t$ is $\Delta t/\tau$.  If a particle scatters,
it is reassigned a temperature and horizontal and vertical velocities
randomly selected from gaussian distributions with dispersions given
by equations (B4a) and (B4b).

The equations for the mean flow are quite similar to the original
formulation.  Indeed, the continuity equation for $\rhob$ and the
moment equation for $\vb_z$ remain unchanged from Paper I.  The
temperature equations is $$\eqalignno{
{D\Tb\over{Dt}}&-{\alpha\over{\rhob\cp}}{DP\over{Dt}}
-{1\over{\rhob\cp}}\popz\left(\Kb{\partial\Tb\over\partz}\right)
-{1\over\cp}[A+B(1-\xi^2)\sigma]\sigma^2
-{C\over\cp}\left({\partial\vb\over\partz}\right)^2
-{\Qdb\over{\rhob\cp}}\cr &+{1\over{\rhob\cp}}\popz(\rhob\cp\wtb)
-{\alpha^2\over{\rhob\cp\Tb}}\pPopz\wtb=0.&(B8)\cr}$$ The only
modification is in the term describing the viscous heating resulting
from three-dimensional turbulence.  The mean horizontal velocity
equation is $${D\vb_x\over{Dt}}
+{1\over\rhob}\popz(\rhob\,\overline{wu})=0,\eqno(B9)$$ but
$\overline{wu}$ is clearly zero since there is no preferred horizontal
direction.  Thus, $\vb_x$ is a constant, which we take to be zero.

\centerline{\sectionfont B.2. The $\dot w$, $\dot u$, $\dot\theta$, and
Higher Moment Equations}

We derive the higher moment equations by considering the evolution of
the perturbations $w$, $u$, and $\theta$.  These evolution equations
are $$\eqalignno{\dot w=&-{\alpha\theta\over{\rhob\Tb}}\pPopz
-\left(A+{\partial\vb\over\partz}\right)w
+{1\over\rhob}\popz(\rhob\wwb),&(B10)\cr
\dot u=&-Au,&(B11)\cr
\dot\theta=&\dgt w+{\alpha^2\over{\rhob\cp\Tb}}\pPopz(w\theta-\wtb)-D\theta
+\left({\alpha^2\over{\rhob\cp\Tb}}{DP\over{Dt}}
+{K,_T\over{\rhob\cp}}{\partial^2\over{\partz^2}} +{\dot
Q,_T\over{\rhob\cp}}\right)\theta\cr &+{1\over\rhob}\popz(\rhob\wtb)
+{1\over\cp}(A+B\sigma)(w^2+2u^2-\sigma^2).&(B12)\cr}$$ The essential
modification of the equations from their counterparts in Paper I is
the omission of turbulent viscosity and turbulent diffusion losses.

The second moment equations are:

$w^2$ moment equation:
$${D\wwb\over{Dt}}+{1\over\rhob}\popz(\rhob\wwwb)
+{2\alpha\over{\rhob\Tb}}\pPopz\wtb
+2\left(A+\pvopz+B\sigma\right)\wwb-{2B\xi^2\over 3}\sigma^3=0.
\eqno(B13)$$

$w\theta$ moment equation: $$\eqalignno{
{D\wtb\over{Dt}}&+{1\over\rhob}\popz(\rhob\wwtb)
+{\alpha\over{\rhob\Tb}}\pPopz\ttb+\left(A+D+\pvopz+2B\sigma\right)\wtb
-\dgt\wwb\cr &-\left({\alpha^2\over{\rhob\cp\Tb}}{DP\over{Dt}}
+{K,_T\over{\rhob\cp}}{\partial^2\Tb\over{\partz^2}}
+{Q,_T\over{\rhob\cp}}\right)\wtb
-{\alpha^2\over{\rhob\cp\Tb}}\pPopz\wwtb\cr
&-{1\over\cp}(A+B\sigma)(\wwwb+2\wuub)=0.&(B14)\cr}$$

$\theta^2$ moment equation: $$\eqalignno{
{D\ttb\over{Dt}}&+{1\over\rhob}\popz(\rhob\wttb)
-2\dgt\wtb+2[D+B\sigma(1-\varpi^2)]\ttb\cr
&-2\left({\alpha^2\over{\rhob\cp\Tb}}{DP\over{Dt}}
+{K,_T\over{\rhob\cp}}{\partial^2\Tb\over{\partz^2}}
+{Q,_T\over{\rhob\cp}}\right)\ttb-2{\alpha^2\over{\rhob\cp\Tb}}\pPopz\wttb\cr
&-{2\over\cp}(A+B\sigma)(\wwtb+2\uutb)=0.&(B15)\cr}$$

$u^2$ moment equation:
$${D\uub\over{Dt}}+{1\over\rhob}\popz(\rhob\wuub)
+2(A+B\sigma)\uub-{2B\xi^2\over 3}\sigma^3=0.\eqno(B16)$$

Since there is no preferred horizontal direction, the remaining second
moments, $\wub$ and $\overline{u\theta}$, are zero in steady state, as
are all moments that are odd in $u$, and hence we do not bother to
write these two moment equations.  In the limit that $\xi=\varpi=0$,
these equations reduce to exactly the second moment equations of the
original theory, verifying that the two alternate treatments of
turbulent losses are equivalent at the local level.  In the more
general case that $\xi$ and $\varpi$ are nonzero, the horizontal
velocity dispersion is nonzero and the equations are modified to
accommodate three-dimensional turbulence.

The third moment equations are:

$w^3$ moment equation:
$${D\wwwb\over{Dt}}+{1\over\rhob}\popz(\rhob\wwwwb)
+{3\alpha\over{\rhob\Tb}}\pPopz\wwtb
+\left(3A+3\pvopz+2B\sigma\right)\wwwb
-{3\wwb\over\rhob}\popz(\rhob\wwb)=0.\eqno(B17)$$

$w^2\theta$ moment equation: $$\eqalignno{
{D\wwtb\over{Dt}}&+{1\over\rhob}\popz(\rhob\wwwtb)
+{2\alpha\over{\rhob\Tb}}\pPopz\wttb
+\left(2A+D+2\pvopz+2B\sigma\right)\wwtb
-{2\wtb\over\rhob}\popz(\rhob\wwb)\cr &
-\dgt\wwwb-\left({\alpha^2\over{\rhob\cp\Tb}}{DP\over{Dt}}
+{K,_T\over{\rhob\cp}}{\partial^2\Tb\over{\partz^2}}
+{Q,_T\over{\rhob\cp}}\right)\wwtb -{\alpha^2\over{\rhob\cp\Tb}}\pPopz
(\wwwtb-\wwb\,\wtb)\cr &-{\wwb\over\rhob}\popz(\rhob\wtb)
-{1\over\cp}(A+B\sigma)(\wwwwb+2\wwuub-\wwb\sigma^2)=0.&(B18)\cr}$$

$w\theta^2$ moment equation: $$\eqalignno{
{D\wttb\over{Dt}}&+{1\over\rhob}\popz(\rhob\wwttb)
+{\alpha\over{\rhob\Tb}}\pPopz\tttb
+\left(A+2D+\pvopz+2B\sigma\right)\wttb-{\ttb\over\rhob}\popz(\rhob\wwb)\cr
&-2\dgt\wwtb-2\left({\alpha^2\over{\rhob\cp\Tb}}{DP\over{Dt}}
+{K,_T\over{\rhob\cp}}{\partial^2\Tb\over{\partz^2}}
+{Q,_T\over{\rhob\cp}}\right)\wttb\cr
&-2{\alpha^2\over{\rhob\cp\Tb}}\pPopz(\wwttb-\wtb^2)
-{2\wtb\over\rhob}\popz(\rhob\wtb)\cr
&-{2\over\cp}(A+B\sigma)(\wwwtb+2\wuutb-\wtb\sigma^2)=0.&(B19)\cr}$$

$\theta^3$ moment equation: $$\eqalignno{
{D\tttb\over{Dt}}&+{1\over\rhob}\popz(\rhob\wtttb)
-3\dgt\wttb-3\left({\alpha^2\over{\rhob\cp\Tb}}{DP\over{Dt}}
+{K,_T\over{\rhob\cp}}{\partial^2\Tb\over{\partz^2}}
+{Q,_T\over{\rhob\cp}}\right)\tttb\cr
&-{3\alpha^2\over{\rhob\cp\Tb}}\pPopz(\wtttb-\wtb\,\ttb)
+(3D+2B\sigma)\tttb-{3\ttb\over\rhob}\popz(\rhob\wtb)\cr
&-{3\over\cp}(A+B\sigma)(\wwttb+2\uuttb-\ttb\sigma^2)=0.&(B20)\cr}$$

$wu^2$ moment equation: $$\eqalignno{
{D\wuub\over{Dt}}&+{1\over\rhob}\popz(\rhob\wwuub)
+{\alpha\over{\rhob\Tb}}\pPopz\wwtb+\left(3A+\pvopz+2B\sigma\right)\wuub\cr
&-{\uub\over\rhob}\popz(\rhob\wwb)=0.&(B21)\cr}$$

$u^2\theta$ moment equation: $$\eqalignno{
{D\uutb\over{Dt}}&+{1\over\rhob}\popz(\rhob\wuutb)
+(2A+D+2B\sigma)\uutb-\dgt\wuub\cr
&-\left({\alpha^2\over{\rhob\cp\Tb}}{DP\over{Dt}}
+{K,_T\over{\rhob\cp}}{\partial^2\Tb\over{\partz^2}}
+{Q,_T\over{\rhob\cp}}\right)\uutb
-{\alpha^2\over{\rhob\cp\Tb}}\pPopz(\wuutb-\wtb\,\uub)\cr
&-{\uub\over\rhob}\popz(\rhob\wtb)
-{1\over\cp}(A+B\sigma)(\wwuub+2\uuuub-\uub\sigma^2)=0.&(B22)\cr}$$

The remaining third moments, $\uuub$, $\wwub$, and $\uttb$, are zero
in steady state, so we have not bothered to write their moment
equations.  These third moment equations are nearly identical to the
original ones, except for the modifications for three-dimensional
motions and one other significant difference.  Because of the form of
the scattering term, turbulent losses enter all these equations (and
indeed would enter all higher moment equations) as a term proportional
to $2B\sigma$.  For example, the damping term of the $w^3$ equation is
$(3A+2B\sigma)\wwwb$.  In the original formulation, turbulent damping
enters all third moments as $3B\sigma$, and, in general, the
coefficient corresponds to the order of the equations.  Although the
two formulations describe the same local behavior in the limit
$\xi=\varpi=0$, the two formulations do not describe the same nonlocal
behavior in any limit.  The differences between these and the original
equations seem minor, but the quantitative effects are significant.

We have written the moment equations in complete detail for comparison
with those in Paper I.  The equations we actually solve in this paper
have had the same simplifications applied as those that yielded the
equations in Appendix A.  Namely, we assume $\vb_z=0$, since this is
true in the steady state, and we neglect $K,_T$ and $\dot Q,_T$ terms,
and terms with $DP/Dt$.

\centerline{\sectionfont B.3. The Local Limit of the Scattering Theory}

In Paper I we showed that if $\xi=0$, $\varpi=0$, in the local limit
these equations are identical to those of the original theory.  Here
we present results for the more general case of nonzero $\xi$ and
$\varpi$.  Recalling that the local limit is obtained when all third
and higher order moments are set equal to zero, we can solve the four
second moment equations (B13)--(B16) to yield a relation between the
velocity dispersion and the unstable superadiabatic gradient,
$${(A+D+2B\sigma)(A+B\sigma)[D+B\sigma(1-\varpi^2)][A+B\sigma(1-\xi^2)]
\over{[A+B\sigma+D+B\sigma(1-\varpi^2)][A+B\sigma(1-\xi^2)]
+(B\sigma\xi^2/3)[D+B\sigma(1-\varpi^2)]}}={g\alpha\over\Tb}\dgt.\eqno(B23)$$
If $\dgt$ is stable (\ie, $\dgt<\Tb AD/g\alpha$), the solution is
$\sigma=0$.  In astrophysical convection, turbulent viscosity is
always dominant over microscopic viscosity, so that $B\sigma/A\gg 1$.
In this limit, the ratio of horizontal to vertical velocity dispersion
is $${\uub\over\wwb}={\xi^2/3\over{1-2\xi^2/3}}.\eqno(B24)$$ Thus, if
there is no scattering into the horizontal direction (\ie, $\xi=0$) in
equation (B4a), there is no horizontal velocity dispersion.  Of
course, even if there is no vertical scattering, the vertical velocity
dispersion remains finite since buoyancy also generates vertical
velocities.  If scattering is maximal ($\xi=1$), $\uub=\wwb$ and the
turbulence is isotropic.

The ratio of turbulent to radiative diffusion may be more or less than
unity in stars.  Convection is efficient if $B\sigma/D\gg 1$, and is
inefficient if $B\sigma/D\ll 1$.  In the limit of efficient
convection, the velocity-temperature correlation is
$${\wtb\over{\sigma_w\sigma_\theta}}=\left[{(2-\varpi^2)(1-\xi^2)
+\xi^2(1-\varpi^2)/3\over{2(1-2\xi^2/3)}}\right]^{1/2}.\eqno(B25)$$ If
$\xi=\varpi=0$, this correlation is unity, but approaches zero as
$\xi$ and $\varpi$ go to unity.  In the inefficient limit, the
correlation is $${\wtb\over{\sigma_w\sigma_\theta}}=1,\eqno(B26)$$
independent of $\xi$ and $\varpi$.  In this limit, the Peclet number,
the efficiency measure, is $${B\sigma\over
D}={g\alpha\Tb\dgt\over{D^2(1-\xi^2)}}.\eqno(B27)$$ Thus, for a given
set of parameters in the inefficient regime, as $\xi$ approaches
unity, convection becomes increasingly efficient, and for large $\xi$,
equation (B26) becomes a poor approximation.

\vfill\eject
\references


\rj{Alongi, M.,  Bertelli, G., Bressan, A., \& Chiosi, C. 1991}{\aa}{244}{95}
\rj{Basu, S., Antia, H.M., \& Narasimha, D. 1994}{\mnras}{267}{209}
\rj{Bertelli, G., Bressan, A.G., \& Chiosi. C. 1985}{\aa}{150}{33}
\rj{B\"ohm-Vitense 1958}{Z. Astrophys.}{46}{108}
\rj{Bressan, A.G., Bertelli, C., \& Chiosi, C. 1981}{\aa}{102}{25}
\rj{Canuto, V.M. 1992}{\apj}{392}{218}
\rj{Canuto, V.M. 1993}{\apj}{416}{331}
\rj{Canuto, V.M., Minotti, F., Ronchi, C., and Ympa, R.M. 1994}{JAS}
{51}{1605}
\rj{Cattaneo, F. Brummell, N.H., Toomre, J., Malagoli, A. 1991}{\apj}{370}{282}
\rj{Chin, C.-w. \& Stothers, R.B. 1991}{\apjs}{77}{299}
\rj{Chiosi, C. Bertelli, G. Meylan, G. \& Ortolani, S. 1989}{\aa}{219}{167}
\rj{Doom, C. 1982a}{\aa}{116}{303}
\rj{Doom, C. 1982b}{\aa}{116}{308}
\rj{Grossman, S.A., Narayan, R., \& Arnett, D. 1993}{\apj}{407}{284 (Paper I)}
\rj{Grossman, S.A., \& Narayan, R. 1993}{\apjs}{89}{361 (Paper II)}
\rj{Hossain, M. \& Mullan, D.J. 1991}{\apj}{380}{631}
\rj{Hurlburt, N.E., Toomre, J., \& Massaguer, J.M. 1986}{\apj}{311}{563}
\rj{Hurlburt, N.E., Toomre, J., Massaguer, J.M., \& Zahn, J.-P. 1994}{\apj}
{421}{245}
\rj{Kumar, P, Narayan, R., \& Loeb, A. 1995}{preprint}{}{}
\rj{Langer, N. 1986}{\aa}{164}{45}
\rb{Lesieur, M. 1987}{Turbulence in Fluids}{Boston: Martinus-Nijhoff}
\rj{Maeder, A. 1975}{\aa}{40}{303}
\rj{Maeder, A. \& Mermilliod, J.-C. 1981}{\aa}{93}{136}
\rj{Maeder, A. \& Meynet, G. 1988}{\aas}{76}{411}
\rj{Maeder, A. \& Meynet, G. 1989}{\aa}{210}{155}
\rj{Mermilliod, J.-C. \& Maeder, A. 1986}{\aa}{158}{45}
\rj{Monteiro, M.J.P.F.G., Christensen-Dalsgaard, J., \& Thompson, M.J. 1994}
{\aa}{283}{247}
\rj{Narayan, R., Loeb, A.,\& Kumar, P. 1994}{\apj}{431}{359}
\rc{Orszag, S.A. 1977}{Fluid Dynamics}{R. Balian \& J.-L. Peube}
{New York: Gordon \& Breach}{235}
\rb{Press, W.H., Flannery, B.P., Teukolsky, S.A., \& Vetterling, W.T. 1986}
{Numerical Recipes}{Cambridge: Cambridge Univ. Press}
\rj{Renzini, A. 1987}{\aa}{188}{49}
\rj{Roxburgh, I.W. 1978}{\aa}{65}{281}
\rj{Roxburgh, I.W. 1989}{\aa}{211}{361}
\rj{Roxburgh, I.W. 1992}{\aa}{266}{291}
\rj{Schmitt, J.H.M.M., Rosner, R., \& Bonn, H.U. 1994}{\apj}{282}{316}
\rj{Shaviv, G. \& Salpeter, E.E. 1973}{\apj}{184}{191}
\rj{Singh, H.P., Roxburgh, I.W., Chan, K.L. 1994}{\aa}{281}{L73}
\rj{Stein, R.F. \& Nordlund, \AA. 1989}{\apj}{342}{L95}
\rj{Stothers, R.B. 1991}{\apj}{383}{820}
\rj{Stothers, R.B. \& Chin, C.-w. 1992}{\apj}{390}{136}
\rj{Travis, L.D. \& Matsushima, S. 1973}{\apj}{180}{975}
\rj{Umezu, M. 1992}{\mnras}{258}{107}
\rj{Unno, W., Kondo, M.-a., \& Xiong, D.-r. 1985}{PASJ}{37}{235}
\rj{Unno, W. \& Kondo, M.-a. 1989}{PASJ}{41}{197}
\rj{Xiong, D.-r. 1980}{Chinese Astron.}{4}{234}
\rj{Xiong, D.-r. 1981}{Sci. Sinica}{24}{1406}
\rj{Xiong, D.-r. 1985}{\aa}{150}{133}
\rj{Xiong, D.-r. 1986}{\aa}{167}{239}
\rj{Xiong, D.-r. 1989}{\aa}{213}{176}
\rj{Xiong, D.-r. 1990}{\aa}{232}{21}
\rj{Xiong, D.-r. \& Chen, Q.L. 1992}{\aa}{254}{362}
\rj{Zahn, J.-P. 1991}{\aa}{252}{179}

\figures

\fig 1.  The superadiabatic gradient and second moments for
an efficient and inefficient overshooting model.  The transition from
instability to stability is discontinuous, as assumed in the analysis
of \S2.2.  The distances for the second moments to decay by a factor
of $e$ beyond the stability transition at $z=0.4$ is comparable to the
distances estimated by equations (6), (7), and (9).  Note that in the
efficient model the local and nonlocal solutions have comparable
amplitude in the convective region because the convective flux is
constrained to be nearly the total flux.  The comparison is not as
favorable in the inefficient case, since there is no strong constraint
on the convective flux.  Here and throughout this paper, we take
increasing $z$ and positive velocities in the direction opposing
gravity.

\fig 2.  The superadiabatic gradient and the three second moments for
the efficient and inefficient homogeneous convection models.  Shown
are the velocity and temperature dispersions, $\sigma_w$ and
$\sigma_\theta$, and the velocity-temperature correlation,
$\wtb_n=\wtb/\sigma_w\sigma_\theta$.  The brackets on $w\theta$
indicate the same ensemble average as overbars in the text, and the
subscript $n$ indicates that $w\theta$ has been normalized.  The Full
solutions have a more local character than the Xiong solutions.

\fig 3.  The four third moments for then efficient and inefficient
homogeneous models.  The Full solutions are better in nearly all
cases.

\fig 4.  The four required fourth moments for the efficient and inefficient
homogeneous models.

\fig 5.  The normalized fourth moments for the efficient and inefficient
homogeneous models.  For reference, the quasi-gaussian (Q.-G.) closure
is shown.

\fig 6.  The superadiabatic gradient and the three second moments for
the efficient and inefficient overshooting models.  The spikes near
the walls seen in some moments are artifacts of the GSPH algorithm and
are not physical.

\fig 7.  The four third moments for the efficient and inefficient
overshooting models.

\fig 8.  The four required fourth moments for the efficient and inefficient
overshooting models.

\fig 9.  The normalized fourth moments for the efficient and inefficient
overshooting models.  For reference, the quasi-gaussian (Q.-G.)
closure is shown.

\fig 10.  The ballistic evolution of particles overshooting in a fluid
described by the superadiabatic gradient of the Full solution.  Plots
of $\theta$ \vs $w$, $w$ \vs $z$, and $\theta$ \vs $z$ are shown for
both the efficient and inefficient overshooting models.  In each case
two integrations are shown.  The solid curves are for integrations
starting at $z=0.4$, at the stability transition, where $w$ and
$\theta$ have already started to decay.  The dotted curves are for
integrations beginning at $z=0.3$.  These particles actually penetrate
slightly farther in the efficient overshooting model.

\fig 11.  The superadiabatic gradient and four second moments
for the efficient and inefficient homogeneous models.  Both the moment
equations and the GSPH code have been modified to treat turbulent
losses as scatterings of turbulent particles.  Note that both the
vertical and horizontal velocity dispersions, $\sigma_w$ and
$\sigma_u$, are shown in the same panels.  The set of lines for
$\sigma_u$ are always below those for $\sigma_w$.

\fig 12.  The four dominant third moments of the efficient and inefficient
homogeneous models.  Although we solve for $\wuub$ and $\uutb$, we do
not show them to avoid too much confusion.  They are much smaller than
the moments $\wwwb$ and $\wwtb$.

\fig 13.  The four dominant fourth moments of the efficient and inefficient
homogeneous models.  We have not shown the moments $\wwuub$, $\wuutb$,
$\uuuub$, or $\uutb$ to avoid confusion in the figure.

\fig 14.  The normalized fourth moments for the efficient and inefficient
homogeneous models.  For reference, the quasi-gaussian (Q.-G.) closure
is shown.

\fig 15.  The superadiabatic gradient and four second moments
for the efficient and inefficient overshooting models.  We show
$\sigma_w$ and $\sigma_u$ in the same panels.  The set of lines for
$\sigma_u$ is always the lower one.

\fig 16.  The four dominant third moments of the efficient and inefficient
overshooting models.  Again we have omitted the curves of third
moments involving $u$.

\fig 17.  The four dominant fourth moments of the efficient and inefficient
overshooting models.  Again we have omitted the curves of fourth
moments involving $u$.

\fig 18.  The normalized fourth moments for the efficient and inefficient
overshooting models.  For reference, the quasi-gaussian (Q.-G.)
closure is shown.


\bye